\documentstyle[prb,aps,multicol,epsf]{revtex}

\def\gsim{\,$\raise0.3ex\hbox{$>$}\llap{\lower0.8ex\hbox{$\sim$}}$\,}
\def\lsim{\,$\raise0.3ex\hbox{$<$}\llap{\lower0.8ex\hbox{$\sim$}}$\,}

\begin{document}

\draft
\title{Ground State Phase Diagrams of Frustrated Spin-$S$ $XXZ$ Chains: \\
       Chiral Ordered Phases}
\author{T. Hikihara$^{1}$, M. Kaburagi$^{2,3}$ and H. Kawamura$^1$}
\address{$^1$Department of Earth and Space Science, 
Graduate School of Science, Osaka University, \\
Toyonaka, Osaka 560-0043, Japan\\
$^2$Faculty of Cross-Cultural Studies, 
Kobe University, Tsurukabuto, Nada, Kobe 657-8501, Japan\\
$^3$Graduate School of Science and Technology, Kobe University,
Rokkodai,  Kobe 657-8501, 
Japan}
\date{October 19, 2000}
\maketitle
\begin{abstract}
The ground-state phase diagram of the frustrated spin-$S$ $XXZ$ chain 
with the competing nearest- and next-nearest-neighbor antiferromagnetic 
couplings is studied numerically 
by using the density-matrix renormalization-group method 
for the cases of $S = 1/2$, $3/2$, and $2$.
We are paricularly interested in 
the possible gapless and gapped chiral phases, 
in which the chirality $\kappa_l = S_l^x S_{l+1}^y - S_l^y S_{l+1}^x$ 
exhibits a finite long-range order whereas the spin correlation 
decays either algebraically or exponentially.
We show that the gapless chiral phase appears in a broad region 
of the phase diagram for general $S$.
By contrast, the gapped chiral phase is found for integer $S$ 
in a narrow region of the phase diagram, 
while it has not been identified for half-odd integer $S$
within our numerical accuracy.
By combining the results with our previous result for $S = 1$, 
we discuss the $S$-dependence of the phase diagram.
The prediction from a bosonization analysis on the decay exponent 
of the spin correlation is verified.

\end{abstract}

\begin{multicols}{2}

\section{Introduction}

The study of frustrated quantum spin chains has been the subject of 
much interest for many years.
This is mainly because these systems exhibit a wide variety of exotic phases 
accompanied with various types of spontaneous symmetry breaking 
due to the interplay between frustration and quantum fluctuations.
Among them, one of the simplest model may be the quantum spin chain 
with the antiferromagnetic (AF) nearest-neighbor coupling $J_1$ 
and the frustrating AF next-nearest-neighbor coupling $J_2$.
The model Hamiltonian has the form, 
\begin{equation}
{\cal H} = \sum_{\rho=1}^{2}\left\{ J_{\rho}\sum_{l} 
           \left(S_l^x S_{l+\rho}^x + S_l^y S_{l+\rho}^y
           + \Delta S_l^z S_{l+\rho}^z \right)\right\}, 
\label{eq:Ham}
\end{equation}
where $\vec{S}_l$ is the spin-$S$ operator at site $l$ and 
$\Delta$ is the exchange anisotropy.
Throughout this paper, we consider the case of 
$0 \le \Delta \le 1$ and $j \equiv J_2/J_1 > 0$.

In the classical limit $S \to \infty$, the spin chain (\ref{eq:Ham}) 
exhibits a magnetic long-range order (LRO) in the ground state
characterized by a certain wavenumber $q$.
The order parameter is defined by 
\begin{equation}
\vec{m}(q) = \frac{1}{LS} \sum_{l} \vec{S}_l e^{iql}, \label{eq:hel}
\end{equation}
where $L$ is the total number of spins.
The LRO is of the N{\`e}el-type ($q = \pi$)
when the frustration is smaller than a critical value, $j \le 1/4$, 
whereas it becomes of helical-type for $j > 1/4$ 
with a wavenumber $q = \cos^{-1} \left( -1/4j \right) $.
Both the time-reversal and parity symmetries 
are broken in this helical ordered phase.
In the $XY$-like case ($0\leq \Delta < 1$), the helical ordered state 
possesses a two-fold discrete chiral degeneracy characterized 
by the right- and left-handed chirality, 
in addition to a continuous degeneracy associated with 
the original $U(1)$ symmetry of the $XY$ spin.
The chiral order parameter is defined by~\cite{chlOP}
\begin{eqnarray}
O_\kappa &=& \frac{1}{LS^2} \sum_{l} \kappa_{l},\label{eq:chl} \\
\kappa_{l} &=& S_{l}^x S_{l+1}^y - S_{l}^y S_{l+1}^x
          = \left[ \vec{S}_{l} \times \vec{S}_{l+1} \right]_z.  \nonumber
\end{eqnarray}
Note that this vector chirality $\kappa_l$ is distinct 
from the scalar chirality often discussed in the literature\cite{Frahm} 
defined by 
$\chi _l= \vec{S}_{l-1}\cdot \vec{S}_l\times \vec{S}_{l+1}$: 
The chirality $O_\kappa$ changes its sign under the parity operation 
but is invariant under the time-reversal operation, while the scalar 
chirality changes its sign under the both operations.

In the quantum case $S < \infty$, it seems well established 
that no magnetic LRO emerges at least for $0 \le \Delta \le 1$ 
in contrast to the classical limit.
The magnetic LRO (\ref{eq:hel}) is destroyed by quantum fluctuations.
We note that the absence of magnetic LRO in the quantum spin chain 
(\ref{eq:Ham}) is proved rigorously for the $XY$ ($\Delta = 0$) 
and the Heisenberg ($\Delta = 1$) cases.~\cite{Momoi}
By contrast, no theorem prohibiting the spontaneous 
breaking of the parity symmetry has been known.
Accordingly, there remains a possibility of 
the appearance of a novel ^^ ^^ chiral" ordered phase in which 
only the chirality (\ref{eq:chl}) exhibits a LRO 
without the magnetic helical LRO (\ref{eq:hel}).
This phase breaks only the parity symmetry spontaneously 
with preserving both the time-reversal and translational symmetries.

For the $S=1/2$ chain, Nersesyan {\it et al.} predicted, 
using the bosonization technique combined with a mean-field analysis, 
that in the $XY$ case ($\Delta = 0$) the system might exhibit 
a chiral ordered phase with gapless excitations for large $j$.~\cite{Ne}
This gapless chiral phase, however, has not been identified in 
our previous numerical work in which the Binder parameter 
of the chirality was calculated numerically for the $S = 1/2$ $XY$ chain 
with up to $L = 20$ sites 
using the exact-diagonalization (ED) method.~\cite{KKH}
Although Aligia {\it et al.\/}~\cite{ABE} pointed out 
that the system size $L = 20$ might be insufficient 
to deal with the chirality in the large $j$ region 
where the gapless chiral phase was expected, 
the question whether the chiral phase is realized 
in the $S = 1/2$ case has not been clarified so far.

Meanwhile, the situation seems less controversial in the $S = 1$ case.
In the previous works, we determined the ground-state phase diagram 
of the $S = 1$ $XXZ$ chain by means of both the ED and 
the density-matrix renormalization-group (DMRG) calculations, 
and showed that the gapless chiral phase appeared 
in a broad region of the $j$-$\Delta$ plane.~\cite{KKH,HKKT}
This observation was supported by the recent field-theoretical analysis 
by Kolezhuk,~\cite{Kole} in which he obtained the gapless chiral phase 
not only for $S = 1$ but also for general integer $S$.
Lecheminant {\it et al.} extended 
the bosonization analysis by Nersesyan {\it et al.} to general $S$ 
and concluded the appearance of the gapless chiral phase.~\cite{LJA}

An interesting observation which was first revealed 
by our numerical study of the $S = 1$ chain is that 
{\it there exist two different types of chiral phases}.~\cite{KKH,HKKT}
In the one of them, the gapless chiral phase, 
the chiral LRO exists while the string and spin correlations 
decay algebraically suggesting gapless excitations.
In the other, called the gapped chiral phase or 
the ^^ ^^ chiral Haldane" phase, 
the chiral and string LRO's coexist and 
the spin correlation decays exponentially suggesting a finite energy gap. 
The chiral Haldane phase exists in a very narrow but finite region 
between the Haldane and the gapless chiral phases.
For example, in the pure $XY$ case ($\Delta = 0$), 
the system undergoes two successive transitions with increasing $j$, 
first at $j = j_{c1}^{(1)} \simeq 0.473$ from the Haldane phase 
to the chiral Haldane phase, 
and then, at $j = j_{c2}^{(1)} \simeq 0.490$ from the chiral Haldane phase 
to the gapless chiral phase.
The next question which arises naturally is 
whether the gapped chiral phase is even realized for general $S \ne 1$, 
or it is specific to $S = 1$.
In Ref. \onlinecite{Kole}, Kolezhuk suggested that 
the gapped chiral phase existed also for general integer $S$, 
although the method used was not applicable to half-odd integer $S$.
Lecheminant {\it et al.} showed within the bosonization analysis that 
the gapless chiral phase realized at larger $j$ might undergo a transition 
into the gapped chiral phase with decreasing $j$ for any $S$ 
including both integer and half-odd integer $S$.~\cite{LJA}
However, both of these analyses were based on some approximations, 
and the question whether the gapped chiral phase 
exists for general $S \ne 1$ still remains open.

The aim of this paper is to examine whether the gapless and gapped 
chiral phases exist in the frustrated $XXZ$ chain 
(\ref{eq:Ham}) for general $S$.
Furthermore, we wish to clarify how the chiral ordered phases 
change their character as the spin quantum number $S$ increases, 
to be connected to the helical ordered phase 
realized in the classical limit $S \to \infty$.
For these purposes, we numerically determine the ground-state phase diagrams 
of the spin chain (\ref{eq:Ham}) in the cases of $S = 1/2$, $3/2$, and $2$.
The method used is the same as that in our previous work.~\cite{HKKT}
Using the DMRG method, we calculate
appropriate correlation functions associated with 
the order parameters characterizing each phase, and
analyze their long-distance behaviors.
By combining the obtained results with those of the $S = 1$ chain 
determined previously,~\cite{HKKT} we find that 
the gapless chiral phase appears in the cases of general $S \ge 1/2$.
By contrast, the gapped chiral phase 
has been identified only in the cases of integer $S$ ($S = 1$ and $2$): 
It has not been identified within our present numerical accuracy 
in the cases of half-odd integer $S$ ($S = 1/2$ and $3/2$).
It also turns out that the region of the gapless chiral phase 
becomes broader as $S$ becomes larger, 
smoothly converging to the region of the helical ordered phase
in the $S \to \infty$ limit realized at $j \ge 1/4$.

The plan of this paper is as follows.
In Sec. II, we explain the method used in this paper. 
Various correlation functions associated with each order parameter 
are introduced.
We show the results of our DMRG calculation in Sec. III.
The phase diagrams obtained for $S = 1/2$, $3/2$, and $2$ 
are presented in Sec. III A, B, and C, respectively.
By combining the results with the $S = 1$ phase diagram 
determined previously,~\cite{HKKT}
we discuss the $S$-dependence of the phase diagram 
in Sec. IV.
We also compare our result on the decay exponent of the spin correlation 
with the prediction from the bosonization studies.~\cite{Ne,LJA}
Finally, our results are summarized in Sec. V.

\section{Correlation Functions and Numerical Method}
In this section, we introduce various correlation functions 
and explain our numerical method.

For half-odd integer $S$, 
it has been known via previous studies that there exist two distinct phases, 
the spin fluid (SF) and the dimer phases.~\cite{dim} 
The SF phase is characterized 
by gapless excitations above the singlet ground state 
and an algebraic decay of spin correlations.
By contrast, the dimer phase is characterized 
by a finite energy gap above the doubly degenerate ground states 
and an exponential decay of spin correlations.
In the dimer phase, both the parity and translational
symmetries are broken spontaneously.
The order parameter characterizing the dimer phase is given as
\begin{eqnarray}
O_{\rm dim}^{\alpha} &=& \frac{1}{LS^2} \sum_l \tau_l^{\alpha},
~~~~(\alpha = x,z)\label{eq:dim} \\
\tau_l^{\alpha} &=& (-1)^l S_l^{\alpha}S_{l+1}^{\alpha}. 
\nonumber
\end{eqnarray}
For the $S = 1/2$ and $S = 3/2$ open chains, 
we calculate the chiral, dimer, and spin correlation functions 
defined by 
\begin{eqnarray}
C_\kappa (r) &=& \frac{1}{S^4} 
                 \langle \kappa_{l_0-r/2} \kappa_{l_0+r/2} \rangle,
                   \label{eq:Cchl} \\
C_{\rm dim}^{x}(r) &=& \frac{1}{S^4}
         \langle S_{l_0-r/2}^xS_{l_0-r/2+1}^x  
         \times (  S_{l_0+r/2}^xS_{l_0+r/2+1}^x \nonumber \\
&&                     - S_{l_0+r/2+1}^xS_{l_0+r/2+2}^x ) \rangle,
                   \label{eq:Cdim} \\
C_s^\alpha (r) &=& \frac{1}{S^2}
        \langle S_{l_0-r/2}^\alpha S_{l_0+r/2}^\alpha \rangle 
               ~~~(\alpha = x,z),  \label{eq:Ccor} 
\end{eqnarray}
which are associated with the order parameters 
(\ref{eq:chl}), (\ref{eq:dim}), and (\ref{eq:hel}), respectively.
The site number $l_0$ represents the center position of the open chains, 
i.e., $l_0 = L/2$ for even $r$ and $l_0 = (L+1)/2$ for odd $r$.
We note that the number of site $L$ is set to be even.
The notation $\langle \cdots \rangle $ represents the expectation value
at the lowest energy state in the subspace of $S_{total}^z = 0$.

For integer $S$, on the other hand, 
it has been known that there generally exist the SF and 
the Haldane phases,~\cite{Hal}
except for the $S = 1$ case where only the Haldane phase appears 
for $\Delta \ge 0$.~\cite{HKKT,LJA}
The Haldane phase, in which the spin correlation decays exponentially 
with a finite energy gap above the singlet ground state, 
is characterized by the generalized string order parameter~\cite{str,genestr}
\begin{equation}
O_{\rm str} = \frac{1}{LS} \sum_l \exp \left( 
    \sum_{k=1}^{l-1} i \frac{\pi}{S} S_k^z \right) S_l^z. \label{eq:str}
\end{equation}
In the $S = 1$ case, it has also been found that there exists 
the ^^ ^^ double-Haldane" phase in the region of larger $j$,~\cite{HKKT,DH}
although its existence has not been clear in 
the cases of $S \ge 2$.~\cite{Roth}
In order to identify these possible Haldane and double-Haldane phases 
in the $S = 2$ open chain, 
we calculate the generalized string correlation function, 
\begin{equation}
C_{\rm str} (r) = \frac{1}{S^2}
                  \langle S_{l_0 - r/2}^z \exp \left( 
    \sum_{k=l_0-r/2}^{l_0+r/2-1} i \frac{\pi}{2} S_k^z \right) 
    S_{l_0 + r/2}^z \rangle, \label{eq:Cstr}
\end{equation}
in addition to the chiral correlation function $C_\kappa (r)$ 
and the spin correlation function $C_s^\alpha (r)$ introduced above.

We calculate those correlation functions for various fixed values of 
$\Delta$ ($j$) with varying $j$ ($\Delta$), 
and estimate the transition point $j_c$ ($\Delta_c$) by examining 
the $r$-dependence of the correlation functions at long distances.
We employ the infinite-system DMRG algorithm proposed by White.~\cite{White}
In the calculation of the $S = 2$ chain, in particular, 
the accelerated algorithm proposed 
by Nishino and Okunishi is used.~\cite{PWFRG}
The number of kept states $m$ is up to 
$m = 450$, $m = 260$, and $m = 260$ for the $S = 1/2$, $3/2$, and $2$ cases, 
respectively.
Convergence of the data with respect to $m$ is checked 
by consecutively increasing $m$.
The truncation error of the DMRG calculation 
increases dramatically as $j$ becomes larger.
Accordingly, in the calculation of larger $j$, 
we need to keep more and more states in order to achieve 
the $m$-convergence of the data.
Due to this difficulty, our calculation is limited to rather small $j$, i.e.,
$j \le 1.6$ for $S = 1/2$ and $j \le 1$ for $S = 3/2$ and $2$.
In the infinite-system DMRG algorithm, 
the system size is increased by two at each DMRG step 
and the calculation is continued 
until the $L$-convergence of the data has been attained.
We have performed the calculation of typically 1000 DMRG steps 
(corresponding to the system with $L=2000$ sites) 
by confirming the $L$-convergence.
We can thereby safely avoid the finite-size effect arising from
the incommensurate character of the spin correlation
as pointed out by Aligia {\it et al.}~\cite{ABE}

\section{Numerical Results}
\subsection{Spin-$1/2$ case}

We begin with the $S = 1/2$ case.
The frustrated $XXZ$ spin chain (\ref{eq:Ham}) 
for $S = 1/2$ has been studied extensively so far. 
We first review the known properties of the ground-state phase diagram.
With increasing $j$, the system undergoes a Kosterlitz-Thouless (KT) 
phase transition from the SF phase to the dimer phase 
at $j = j_d^{(1/2)}$.~\cite{Tone-Hara,Oka-Nomu}
The critical value $j_d^{(1/2)}$ has been estimated accurately 
for $0 \le \Delta \le 1$:
It runs from $j_d^{(1/2)} (\Delta = 0) \simeq 0.33$ 
in the $XY$ case to $j_d^{(1/2)} (\Delta = 1) \simeq 0.2411$ 
in the Heisenberg case as $\Delta$ increases.~\cite{Oka-Nomu}
Furthermore, the phase diagram is divided into 
two regions according to the nature of the short-range spin correlation.
The structure factor $S(q)$ has a maximum at $q = \pi$ for $j < j_L^{(1/2)}$ 
whereas the maximum of $S(q)$ occurs at an incommensurate position 
$q < \pi$ for $j > j_L^{(1/2)}$.
The Lifshitz point $j_L^{(1/2)}$ in the Heisenberg case ($\Delta = 1$) 
has been estimated to be $j_L^{(1/2)} \simeq 0.52$.~\cite{Bur}
Meanwhile, the phase diagram in the large $j$ region remains largely unclear. 
In particular, as already mentioned in Sec. I, 
the question whether the chiral phase ever exists for larger $j$ 
still remains controversial.

\begin{figure}
\begin{center}
\noindent
\leavevmode\epsfxsize=75mm
\epsfbox{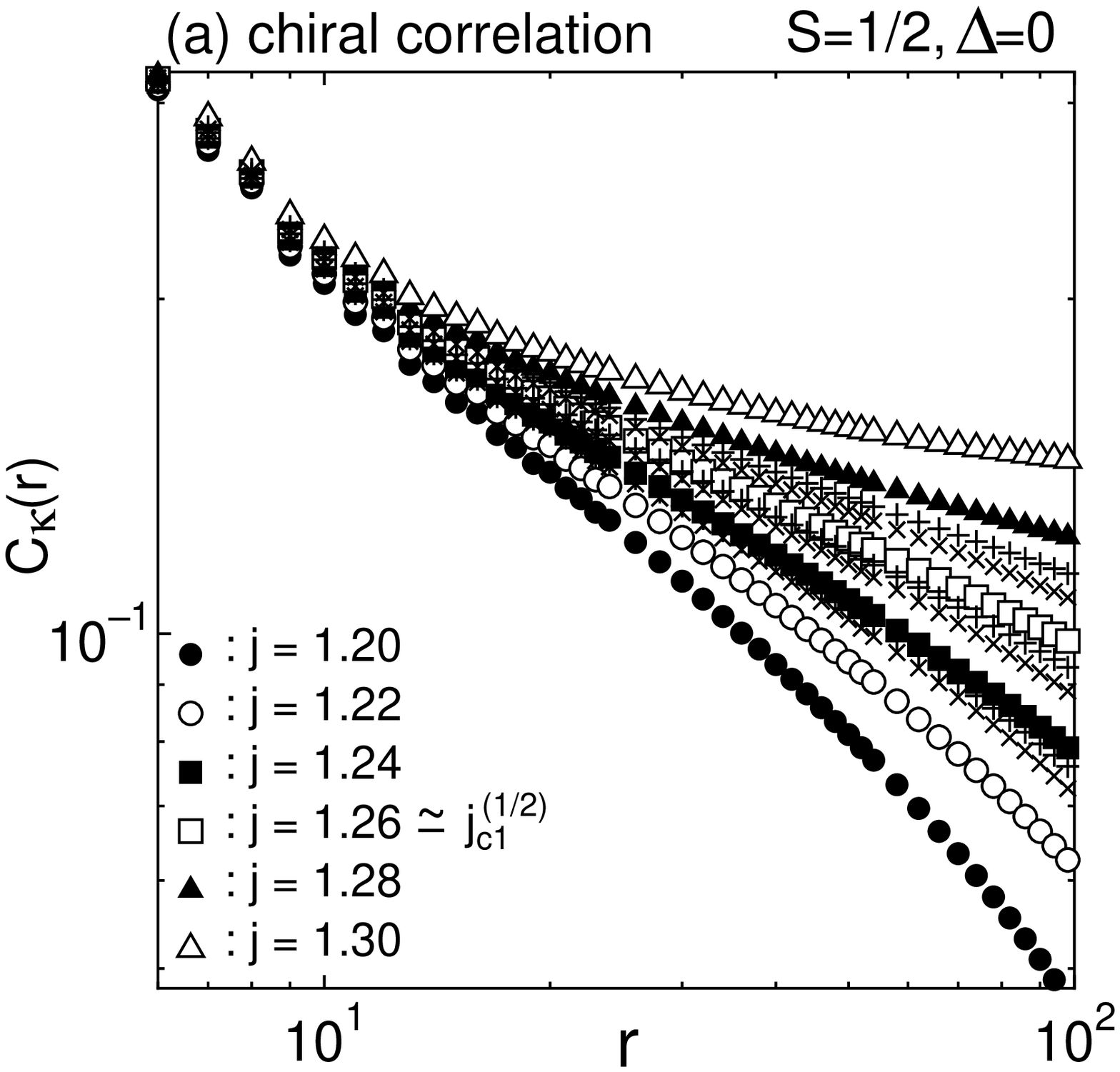}
%\end{center}
%\begin{center}
\noindent
\leavevmode\epsfxsize=75mm
\epsfbox{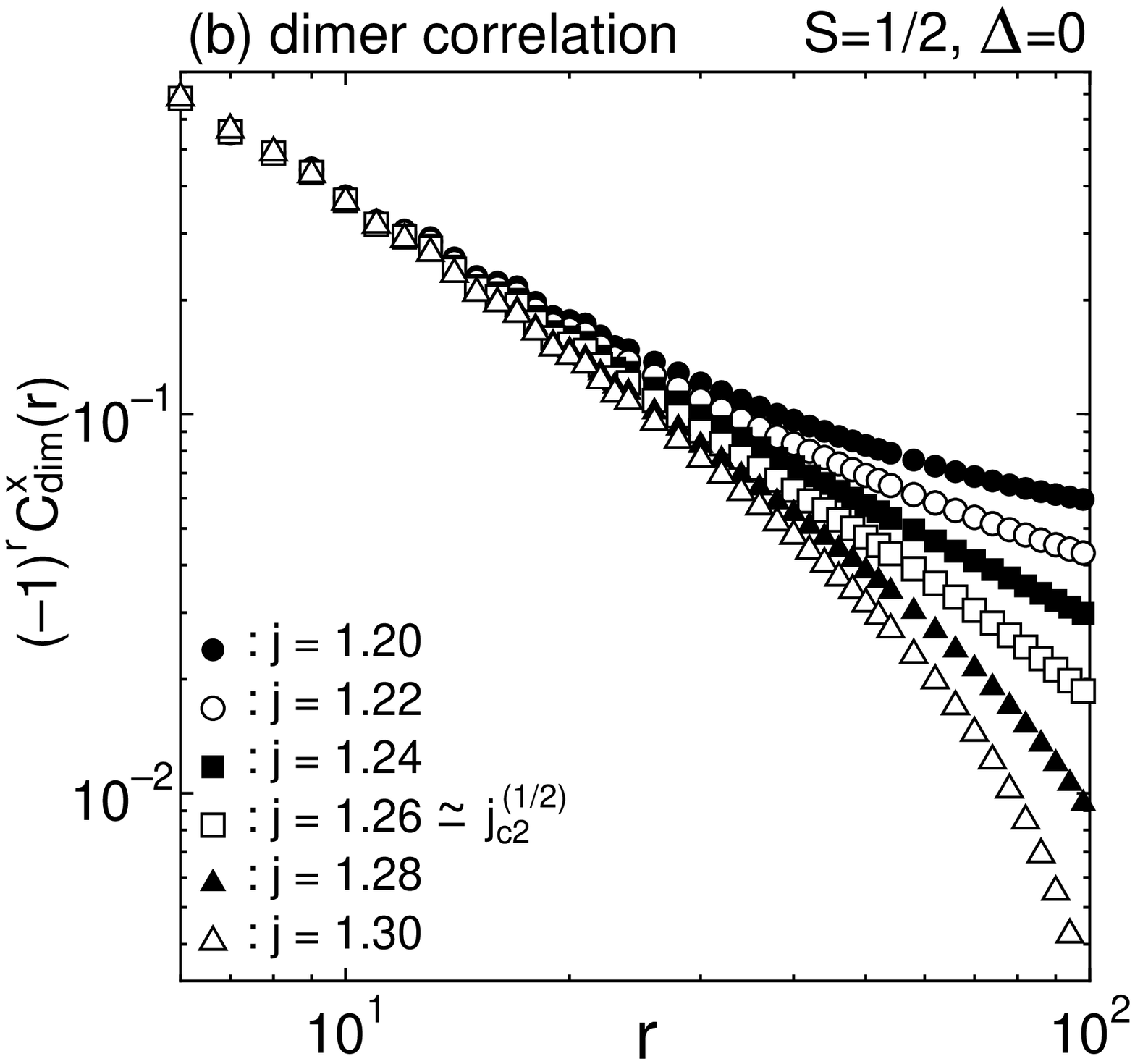}
\end{center}
\begin{center}
\noindent
\leavevmode\epsfxsize=75mm
\epsfbox{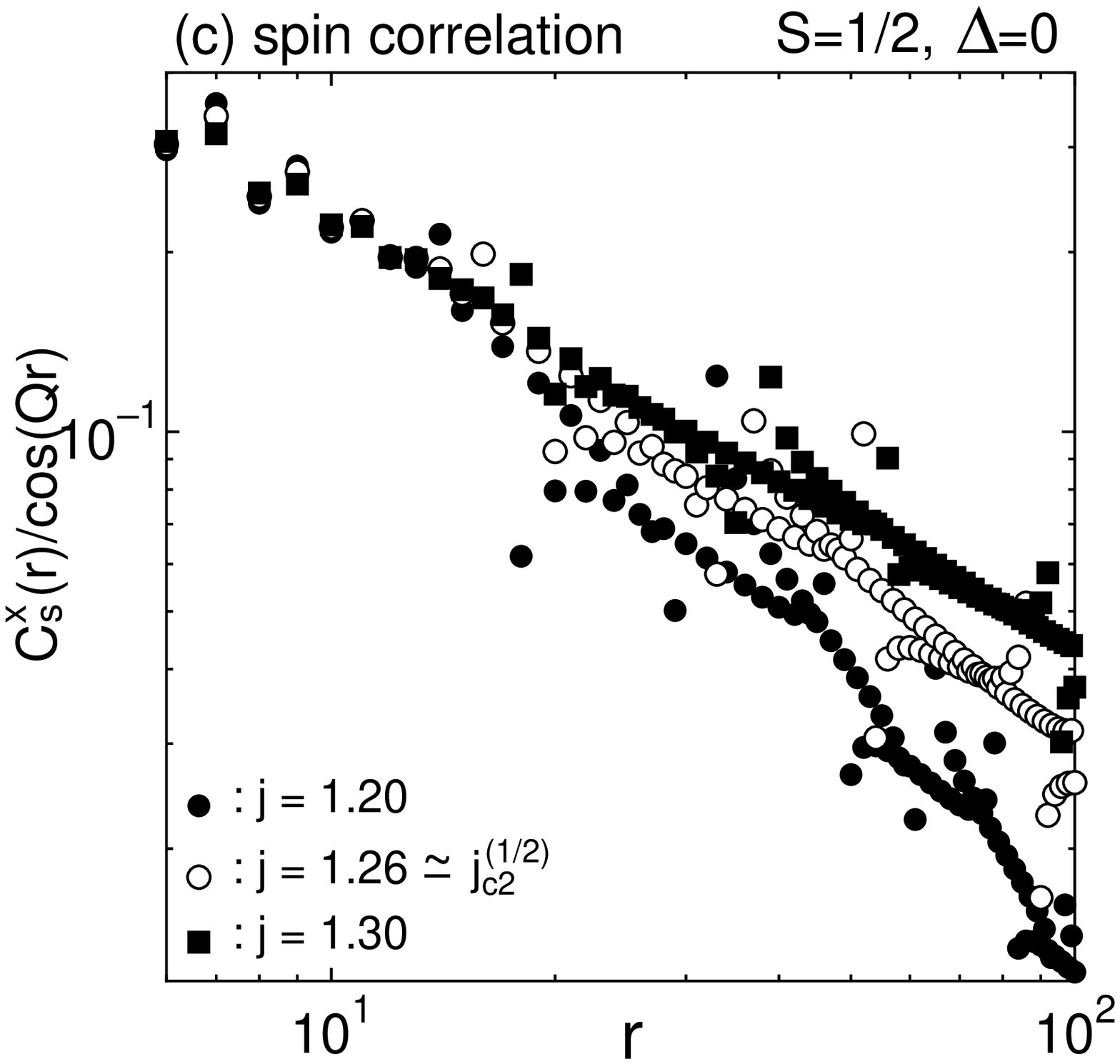}
\end{center}
\caption{Various correlation functions of the $S = 1/2$ chain 
for $\Delta = 0$ and for several typical values of $j$ 
around the dimer-chiral transition: 
(a) chiral correlation function $C_\kappa (r)$;
(b) dimer correlation function $(-1)^r C_{\rm dim}^x(r)$;
(c) spin correlation function $C_{s}^x(r)$ divided by the oscillating factor 
$\cos (Qr)$. 
The number of kept states is $m=450$.
To illustrate the $m$-dependence, we also 
indicate by crosses the data with $m=400$ and $350$ for 
$j = 1.24, 1.26$, and $1.28$ in figure (a) 
where the $m$-dependence is relatively large.
In other cases, the truncation errors are smaller than the symbols.
}
\label{fig:1hD-C}
\end{figure}

Now, we show our numerical results on the dimer-chiral transition.
The calculated chiral, dimer, and spin correlation functions 
are shown in Fig.\ref{fig:1hD-C} (a)-(c) on log-log plots
for $\Delta = 0$ and for several typical values of $j$.
As can clearly be seen from Fig.\ref{fig:1hD-C} (a), 
the chiral correlation function $C_\kappa (r)$ for 
$j > j_{c1}^{(1/2)} \simeq 1.26$ 
is bent upward at larger $r$ suggesting a finite chiral LRO, 
while $C_\kappa (r)$ for $j < j_{c1}^{(1/2)}$ is bent downward 
suggesting an exponential decay of chiral correlations.
Although the data around the transition point suffer from 
the truncation error inherent to the DMRG calculation, 
we can estimate the transition point as 
$j_{c1}^{(1/2)} = 1.26^{+0.01}_{-0.03}$ 
by taking account of the $m$-dependence of the data shown in the figure.
Thus, we conclude that the chiral ordered phase appears 
even in the $S = 1/2$ chain for $j \ge j_{c1}^{(1/2)} \simeq 1.26$.
The result is consistent with 
the prediction of the bosonization study.~\cite{Ne}

Meanwhile, as shown in Fig.\ref{fig:1hD-C} (b), 
the dimer correlation function $C_{\rm dim}^x (r)$ 
for $j < j_{c2}^{(1/2)} \simeq 1.26$ is 
bent upward for larger $r$ suggesting a finite dimer LRO,
whereas it is bent downward for $j > j_{c2}^{(1/2)}$.
We estimate the dimer transition point $j_{c2}^{(1/2)}$ to be 
$j_{c2}^{(1/2)} = 1.26 \pm 0.01$. 
We note that, contrary to the chiral correlation function, 
the $m$-convergence of $C_{\rm dim}^x (r)$ has 
almost been attained at $m = 450$.
Figure \ref{fig:1hD-C} (c) shows 
the spin correlation function $C_s^x(r)$ divided by 
the leading oscillating factor $\cos(Qr)$, 
where $Q$ is the wavenumber characterizing 
the incommensurability of $C_s^x(r)$ in real space.
Here, the $m$-convergence of the data has also almost been attained.
Although the plots are largely scattered, which 
might be attributed to the possible influence of correction terms 
characterized by wavenumbers $Q' \ne Q$, 
it is still clearly visible in the figure that 
the spin correlation is bent downward 
for $j < j_{c2}^{(1/2)}$ suggesting an exponential decay, 
while it exhibits a linear behavior for $j > j_{c2}^{(1/2)}$ 
suggesting a power-law decay.
From these behaviors of the dimer and spin correlations,
we conclude that, as $j$ increases, the system exhibits a transition 
from the gapped phase with the dimer LRO ($j < j_{c2}^{(1/2)}$) 
to the gapless phase without the dimer LRO ($j > j_{c2}^{(1/2)}$).

The remaining problem is the relation 
between $j_{c1}^{(1/2)}$ and $j_{c2}^{(1/2)}$.
Two possibilities seem to be allowed from our data, i.e., 
(i) $j_{c1}^{(1/2)} = j_{c2}^{(1/2)}$ or 
(ii) $j_{c1}^{(1/2)} < j_{c2}^{(1/2)}$.
If the case (i) is realized, the system undergoes only one phase
transition 
at $j = j_{c1}^{(1/2)} = j_{c2}^{(1/2)}$ between the dimer phase 
and the gapless chiral phase with no dimer order.
If the case (ii) is realized, on the other hand, 
the system undergoes two successive transitions on increasing $j$,
first at $j = j_{c1}^{(1/2)}$ from the dimer phase to 
the gapped chiral phase (or the ^^ ^^ chiral dimer" phase) 
where both the dimer and chiral LRO's coexist with gapfull excitations,
and then at $j = j_{c2}^{(1/2)}$ from 
the chiral dimer phase to the gapless chiral phase.
Unfortunately, rather large error bars of 
$j_{c1}^{(1/2)}$ and $j_{c2}^{(1/2)}$ prevent us from determining 
which of the cases is realized.
However, our result suggests that the chiral dimer phase, 
if it ever exists, appears 
only in a rather narrow region, less than about 3\% of $j_c \simeq 1.26$, 
between the dimer and gapless chiral phases.
(We note that in the $S=1$ $XY$ chain the gapped chiral phase 
 exists for $0.473 \lsim j \lsim 0.49$,  whose width corresponds to 
 about 3.6\% of $j_c \simeq 0.473$.~\cite{KKH,HKKT})
Further work will be necessary to settle 
the question whether the chiral dimer phase exists in the $S = 1/2$ case.

Performing the calculations for several fixed values of $\Delta$ ($j$) 
with varying $j$ ($\Delta$), 
we determine the phase boundary between the dimer and 
the gapless chiral phases.
The obtained transition points $j_{c1}^{(1/2)}$ ($\Delta_{c1}^{(1/2)}$) are 
plotted in Fig. \ref{fig:1hdiag} together with the SF-dimer transition points 
$j_d^{(1/2)}$ determined by Okamoto and Nomura.~\cite{Oka-Nomu}
As can be seen from the phase diagram, the gapless chiral phase 
does exist in the large $j$ region.
At least within our present numerical precision, 
on the other hand, we did not find any evidence of the gapped chiral 
(chiral dimer) phase at all estimated points 
between the dimer and the gapless chiral phases.

Finally, we mention the reason why the ED analysis of Ref. \onlinecite{KKH}
on the chain up to $L = 20$ sites failed to detect the chiral ordering 
in the $S = 1/2$ $XY$ chain.
We now consider that this failure can be ascribed to the finite-size effect 
discussed in Ref. \onlinecite{ABE}.
The wavenumber $Q$ characterizing the incommensurability 
decreases rapidly from $Q = \pi$ at $j = j_L$ 
to $Q = \pi/2$ at $j \to \infty$ as $j$ increases.
In the $S=1/2$ $XY$ chain, the shift of $Q$ from $\pi/2$ 
turns out to be smaller than $0.03\pi$ 
for $j \gsim j_{c1}^{(1/2)} \simeq 1.26$.
Meanwhile, in a finite open chain with $L$ sites, 
the incommensurability smaller than $2\pi/L$ cannot be taken into account 
because of the condition that the wavefunction should vanish on both ends 
of the chain.
As a consequence, the system size must be larger than $L \sim 70$ 
if one wishes to detect the shift of $Q$ of order $0.03\pi$.
Thus, the ED calculation on chains up to $L=20$ might fail to 
extract the true asymptotic properties of the chiral Binder parameter 
for $j \gsim j_{c1}^{(1/2)}$.

\subsection{Spin-3/2 case}

In this subsection, we present the results for the $S = 3/2$ case.
The frustrated $S = 3/2$ chain (\ref{eq:Ham}) was studied 
for the Heisenberg case ($\Delta = 1$).~\cite{Roth,Zi-Sch}
It was shown there that the system was in the SF phase for $j < j_d^{(3/2)}$ 
while it was in the dimer phase for $j > j_d^{(3/2)}$.
The phase transition of the KT type occurs at $j = j_d^{(3/2)} \simeq 0.33$.
The Lifshitz point was estimated to be at 
$j = j_L^{(3/2)}(\Delta = 1) = 0.388$.~\cite{Roth}
Besides, the appearance of the gapped and gapless chiral phases 
was suggested recently for the $XY$ case ($\Delta = 0$) 
by the bosonization study.~\cite{LJA}

\begin{figure}
\begin{center}
\noindent
\leavevmode\epsfxsize=75mm
\epsfbox{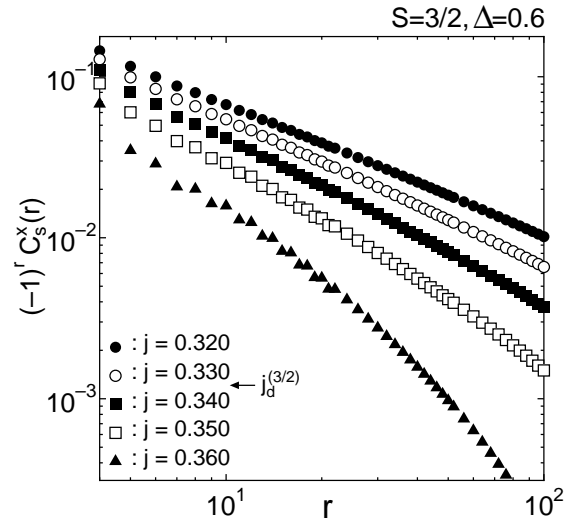}
\end{center}
\caption{The spin correlation function $(-1)^r C_s^x(r)$ 
of the $S = 3/2$ chain 
for $\Delta = 0.6$ and for several typical values of $j$ 
around the SF-dimer transition.
The number of kept states is $m = 220$ for all $j$.
The truncation errors are smaller than the symbols.}
\label{fig:3hSF-D}
\end{figure}

In Fig. \ref{fig:3hSF-D}, 
we show the DMRG data of the spin correlation function $C_s^x (r)$ 
for $\Delta = 0.6$ and for several typical values of $j$.
As can be seen from the figure,
the spin correlation changes its behavior as $j$ increases 
from a power-law decay ($j < j_d^{(3/2)} \simeq 0.335$) 
to an exponential decay ($j > j_d^{(3/2)}$).
We estimate the transition point 
between the gapless SF phase and the gapped dimer phase 
as $j_d^{(3/2)}(\Delta=0.6) = 0.335 \pm 0.015$.
By interpolating the transition points 
estimated in this way for various $\Delta$,
we determine the SF-dimer transition line.
The line runs from $j_d^{(3/2)}(\Delta=0) = 0.334 \pm 0.004$ in the $XY$ case 
to $j_d^{(3/2)}(\Delta=1) = 0.335 \pm 0.015$ in the Heisenberg case, 
the latter being consistent with the estimate by Ziman and Schulz, 
$j_d^{(3/2)}(\Delta=1) \simeq 0.33$.~\cite{Zi-Sch}

\begin{figure}
\begin{center}
\noindent
\leavevmode\epsfxsize=75mm
\epsfbox{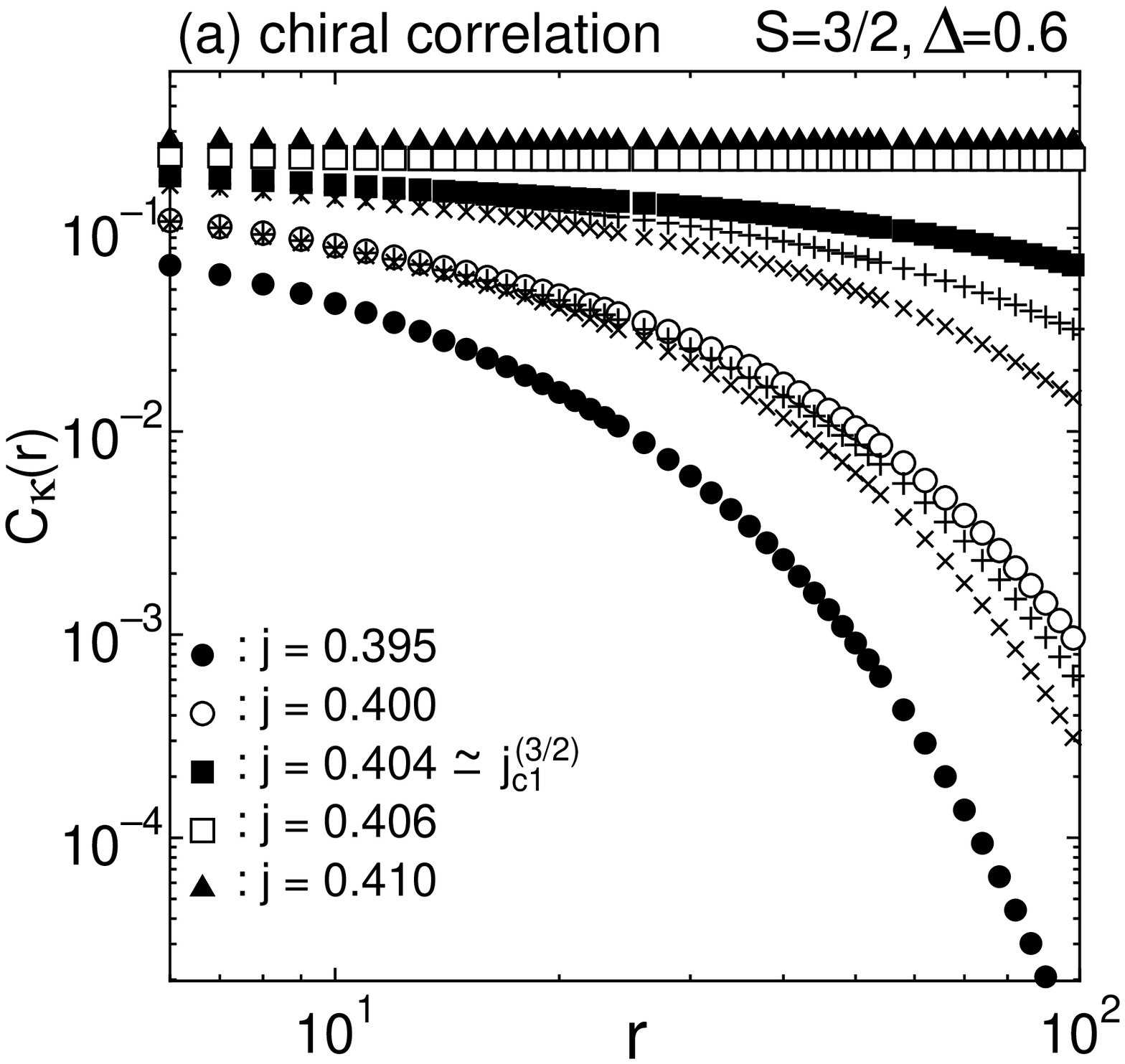}
%\end{center}
%\begin{center}
\noindent
\leavevmode\epsfxsize=75mm
\epsfbox{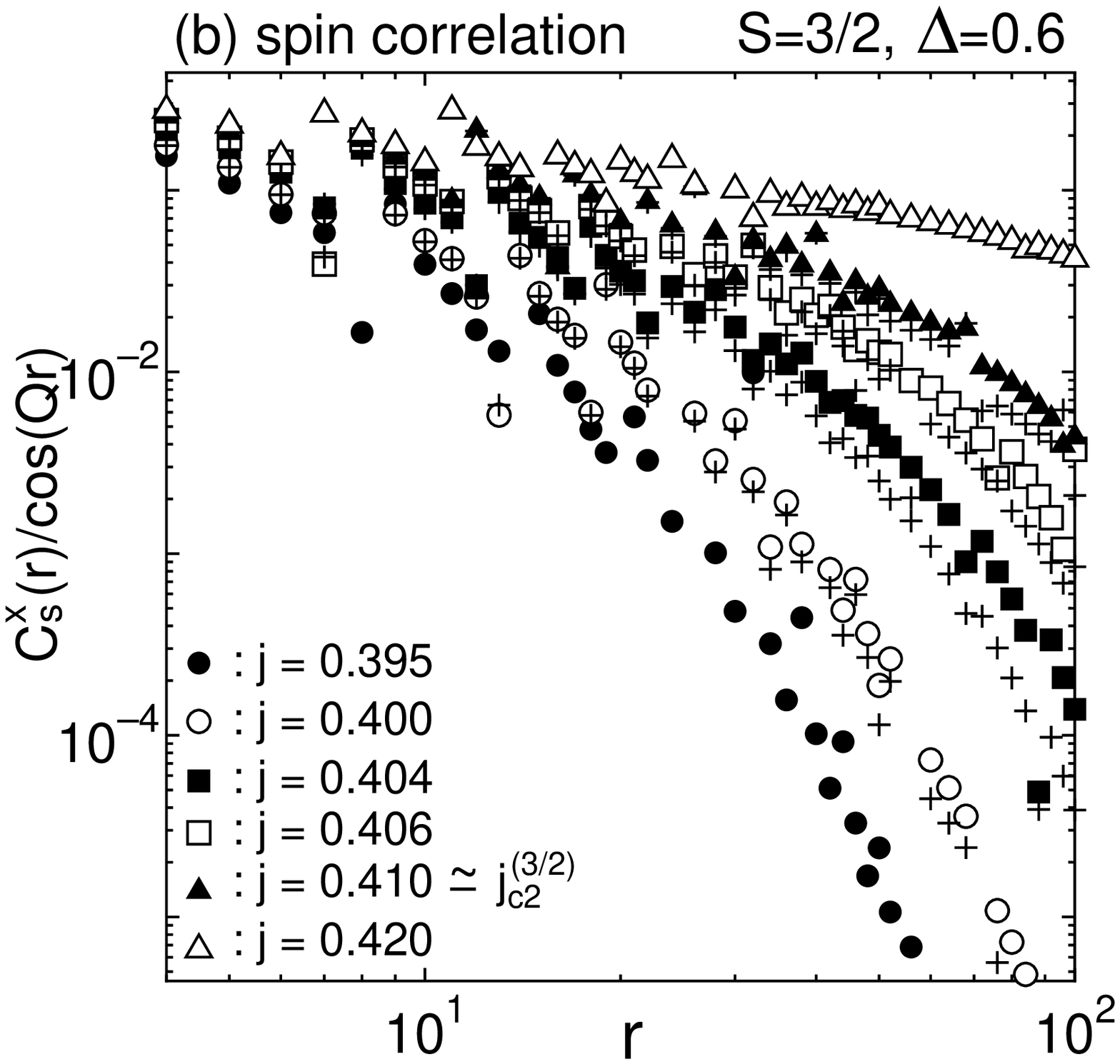}
\end{center}
\caption{Various correlation functions of the $S = 3/2$ chain 
for $\Delta = 0.6$ and for several typical values of $j$ 
around the dimer-chiral transition: 
(a) chiral correlation function $C_\kappa (r)$;
(b) spin correlation function $C_{s}^x(r)$ divided by the oscillating factor 
$\cos (Qr)$. 
The number of kept states is $m = 260$.
We also indicate by crosses the data with $m = 220$ and $180$ 
for $j = 0.400$ and $0.404$ in figure (a), and 
the data with $m = 220$ for $j = 0.400, 0.404, 0.406$, and 
$0.410$ in figure (b).
In other cases, the truncation errors are smaller than the symbols.
}
\label{fig:3hD-C}
\end{figure}

Let us next consider the dimer-chiral transition.
Figure \ref{fig:3hD-C} (a) shows the chiral correlation function 
$C_\kappa (r)$ for $\Delta = 0.6$ and for several typical values of $j$.
As shown in the figure, $C_\kappa (r)$ is bent upward 
for $j > j_{c1}^{(3/2)} \simeq 0.404$ suggesting a finite LRO, 
while it is bent downward for $j < j_{c1}^{(3/2)}$ 
suggesting an exponential decay.
Hence, the chiral ordered phase is realized also in the $S = 3/2$ case.
The transition point at which the chiral LRO sets in is estimated 
to be $j_{c1}^{(3/2)} = 0.404_{-0.006}^{+0.002}$.
The spin correlation function $C_s^x (r)$ for $\Delta = 0.6$ 
is shown in Fig. \ref{fig:3hD-C} (b) where the data are 
divided by the oscillating factor $\cos(Qr)$.
As in the $S = 1/2$ case, the plots are largely scattered 
suggesting the existence of non-negligible correction terms.
Nevertheless, it can be seen that the behavior of $C_s^x (r)$ 
changes as $j$ increases from an exponential decay to an algebraic decay 
at around $j = j_{c2}^{(3/2)} = 0.410 \pm 0.010$.
We thus conclude that the system for $j > j_{c2}^{(3/2)}$ 
is in the gapless chiral phase.
Unfortunately, the error of our estimate of $j_{c2}^{(3/2)}$ is quite large.
This large error is ascribed to the poor $m$-convergence of 
the data of $C_s^x(r)$ shown in Fig. \ref{fig:3hD-C} (b) 
and of the data of the dimer correlation function $C_{\rm dim}^x (r)$ 
to be mentioned below.
(Note that the situation here is different from the $S = 1/2$ case 
where the $m$-convergence of the spin and dimer correlation functions 
is almost achieved at $m = 450$: See Fig. \ref{fig:1hD-C}.)
Because of the errors in the estimates of 
$j_{c1}^{(3/2)}$ and $j_{c2}^{(3/2)}$, 
we can not determine whether $j_{c2}^{(3/2)}$ is either larger 
than or equal to $j_{c1}^{(3/2)}$.
Thus, in our present calculation, 
the gapped chiral phase has not been identified in the $S = 3/2$ case 
as well as in the $S = 1/2$ case.
We have performed the same calculation for various $\Delta$ 
with varying $j$, but have not been able to identify 
the gapped chiral phase at any $j$ and $\Delta$.

\begin{figure}
\begin{center}
\noindent
\leavevmode\epsfxsize=75mm
\epsfbox{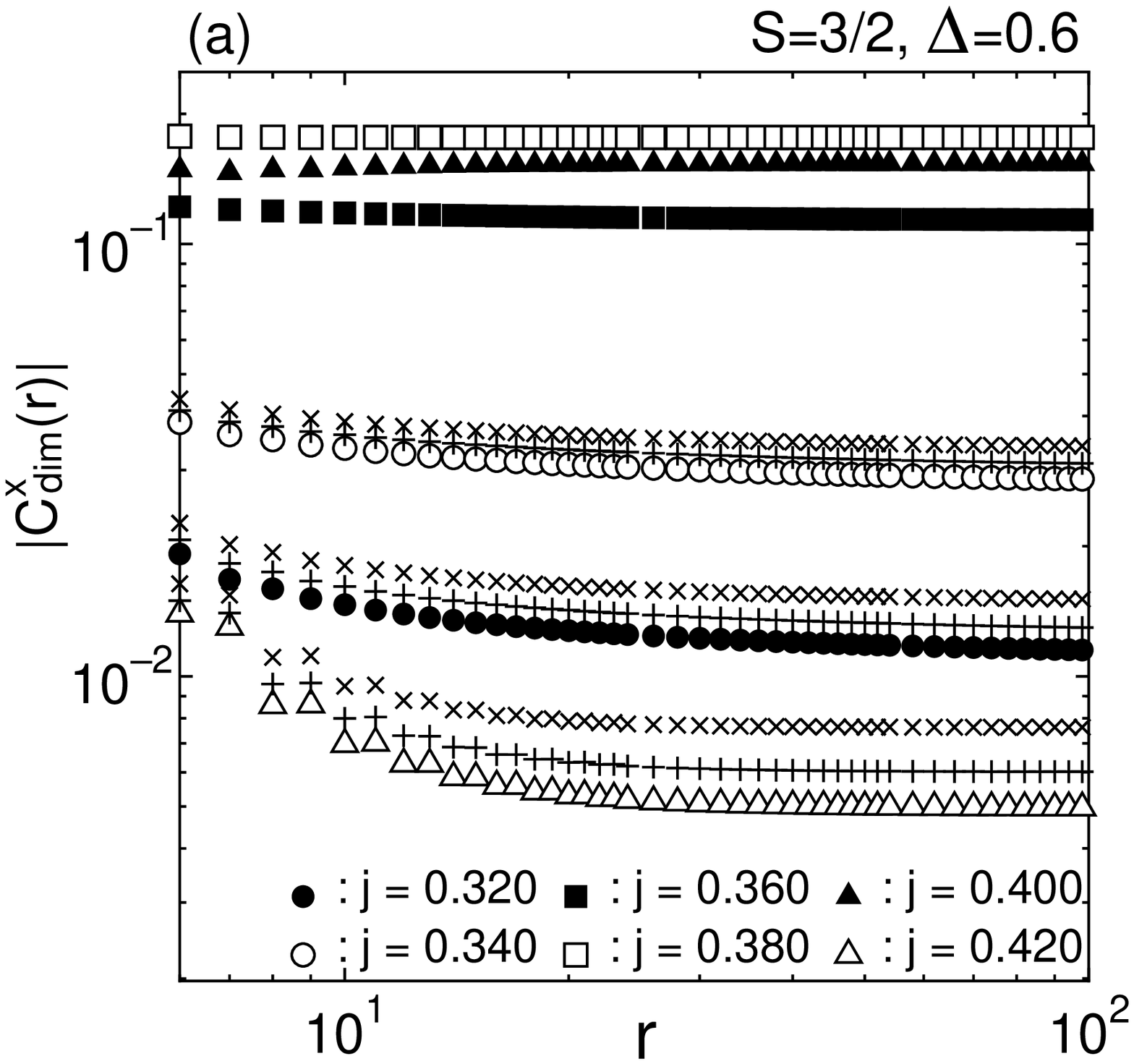}
%\end{center}
%\begin{center}
\noindent
\leavevmode\epsfxsize=75mm
\epsfbox{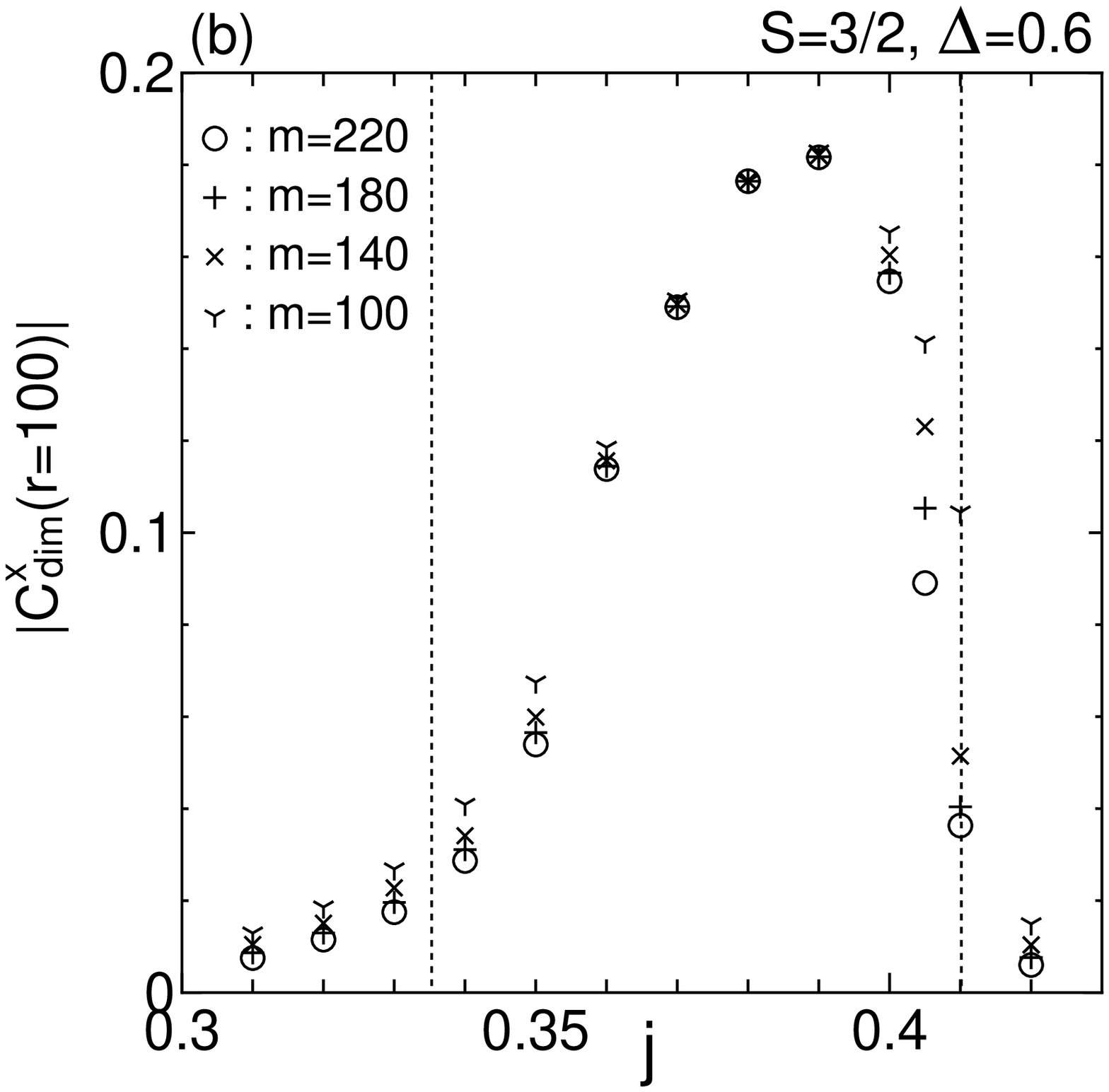}
\end{center}
\caption{(a) The $r$-dependence of the dimer correlation function 
$C_{\rm dim}^x(r)$ of the $S = 3/2$ chain for $\Delta = 0.6$ and 
for several typical values of $j$.
The number of kept states is $m = 220$. 
To illustrate the $m$-dependence of the data, 
we also plot the data with $m = 180$ and $140$ by crosses 
for $j = 0.320, 0.340$, and $0.420$.
The truncation errors for the other $j$'s are smaller than the symbols.
(b) The $j$-dependence of the value of $C_{\rm dim}^x (r = 100)$ 
for various $m$.
The dotted lines represent the phase boundaries, 
$j = j_d^{(3/2)}(\Delta = 0.6) \simeq 0.335$ and 
$j = j_{c2}^{(3/2)}(\Delta = 0.6) \simeq 0.410$.
}
\label{fig:3hCdim}
\end{figure}

In Fig. \ref{fig:3hCdim} (a), we show the dimer correlation function 
$C_{\rm dim}^x (r)$ for $\Delta =0.6$ and for several typical values of $j$.
At a glance, the data may look like indicating a finite dimer LRO even in 
the SF phase ($j < j_d^{(3/2)} \simeq 0.335$) and 
in the gapless chiral phase ($j > j_{c2}^{(3/2)} \simeq 0.410$) 
where no dimer LRO is to be expected.
The dimer correlations in these phases, however, rapidly decrease 
with increasing $m$, whereas the $m$-convergence has almost been attained 
in the intermediate range of $j$ corresponding to the gapped phase 
($j_d^{(3/2)} < j < j_{c2}^{(3/2)}$).
To elucidate the $m$-dependence of the long-distance value of 
the dimer correlation, 
we plot in Fig. \ref{fig:3hCdim} (b) the value of $C_{\rm dim}^x (r=100)$, 
which is expected to be a good approximation of 
$C_{\rm dim}^x (r \to \infty)$.~\cite{Cdim}
The figure shows that $C_{\rm dim}^x (r=100)$ in the gapped phase 
converges to a nonzero value as $m \to \infty$, 
while it decreases toward zero in the SF and the gapless chiral phases.
Thus, we infer that the apparent finite dimer LRO observed 
in $C_{\rm dim}^x (r)$ for $j < j_d^{(3/2)}$ and $j > j_{c2}^{(3/2)}$ 
is a spurious effect due to the truncation error.~\cite{apparent}
The variation of the data with varying $j$ 
near the expected phase boundaries is rather gentle,
which prevents us from estimating the transition points 
accurately from Fig. \ref{fig:3hCdim} (b).
Nevertheless, we may conclude that the system in the gapped region 
$j_d^{(3/2)} < j < j_{c2}^{(3/2)}$ possesses a true dimer LRO 
and is indeed in the dimer phase.

We show in Fig. \ref{fig:3hdiag} the obtained ground-state phase diagram 
of the $S = 3/2$ chain 
including the SF-dimer and the dimer-chiral transition lines, 
$j_d^{(3/2)}$ and $j_{c1}^{(3/2)}$.
We note that the dimer phase exists even in the $XY$ case ($\Delta = 0$) 
for $j_d^{(3/2)} \simeq 0.334 < j < j_{c1}^{(3/2)} \simeq 0.339$
although its width is quite narrow.
The value of $j_{c1}^{(3/2)}$ becomes larger 
as $\Delta$ becomes larger, 
and the dimer-chiral phase boundary appears to tend to 
the Heisenberg line $\Delta = 1$ as $j \to \infty$.

\subsection{Spin-2 case}

In this subsection, we present our results for the $S = 2$ case.
It has been known that the $S = 2$ chain with only the nearest-neighbor 
coupling exhibits a phase transition at 
$\Delta = \Delta_H^{(2)} (j=0) \simeq 0.966$ 
between the SF and the Haldane phases.~\cite{No-Ki}
Meanwhile, the Lifshitz point in the Heisenberg case was estimated 
to be at $j = j_L^{(2)}(\Delta = 1) = 0.325$.~\cite{Roth}
Very recently, the existence of the gapless and gapped chiral phases 
has been suggested by the large-$S$ approach~\cite{Kole} 
and the bosonization method.~\cite{LJA}

\begin{figure}
\begin{center}
\noindent
\leavevmode\epsfxsize=75mm
\epsfbox{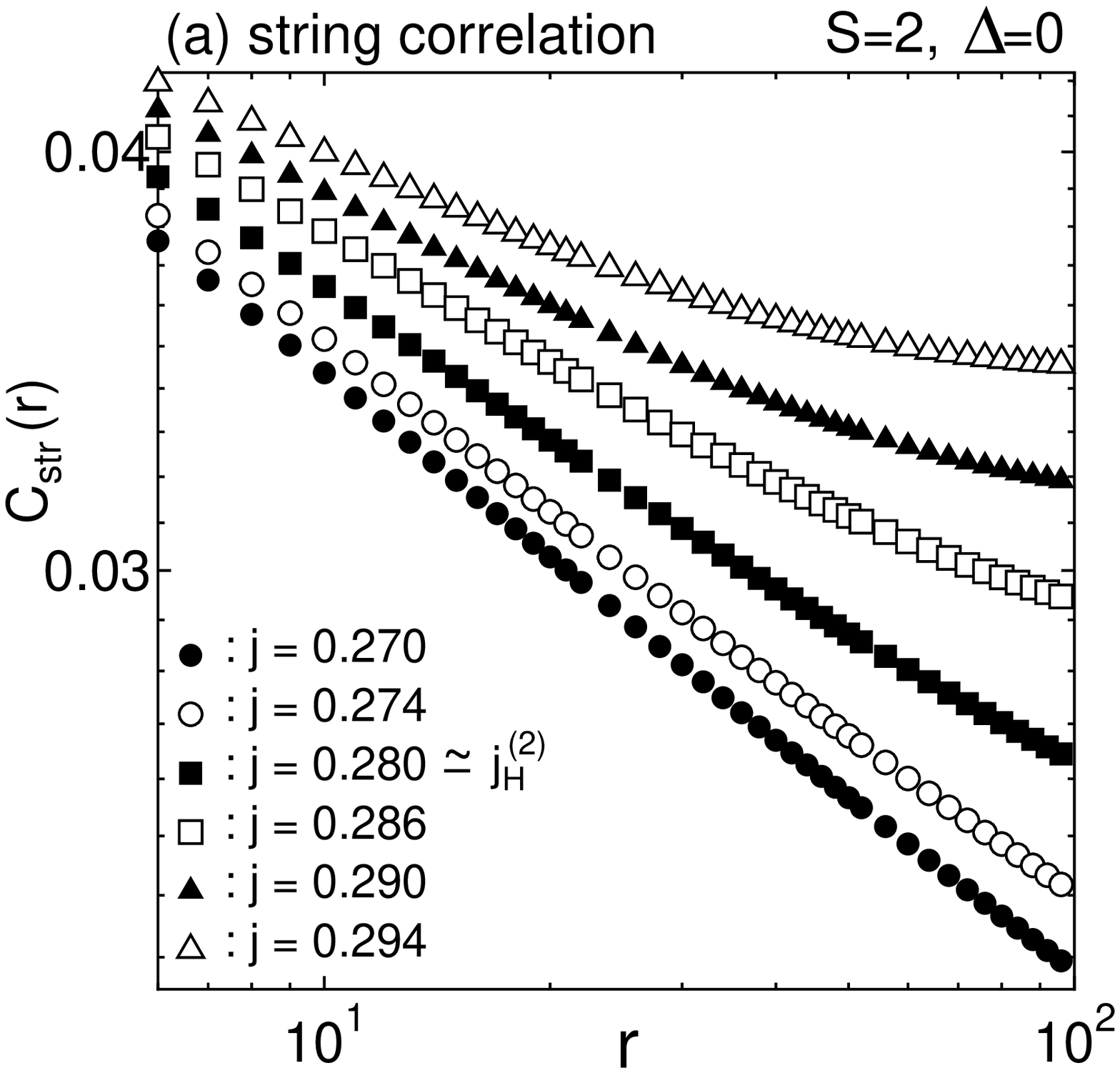}
%\end{center}
%\begin{center}
\noindent
\leavevmode\epsfxsize=75mm
\epsfbox{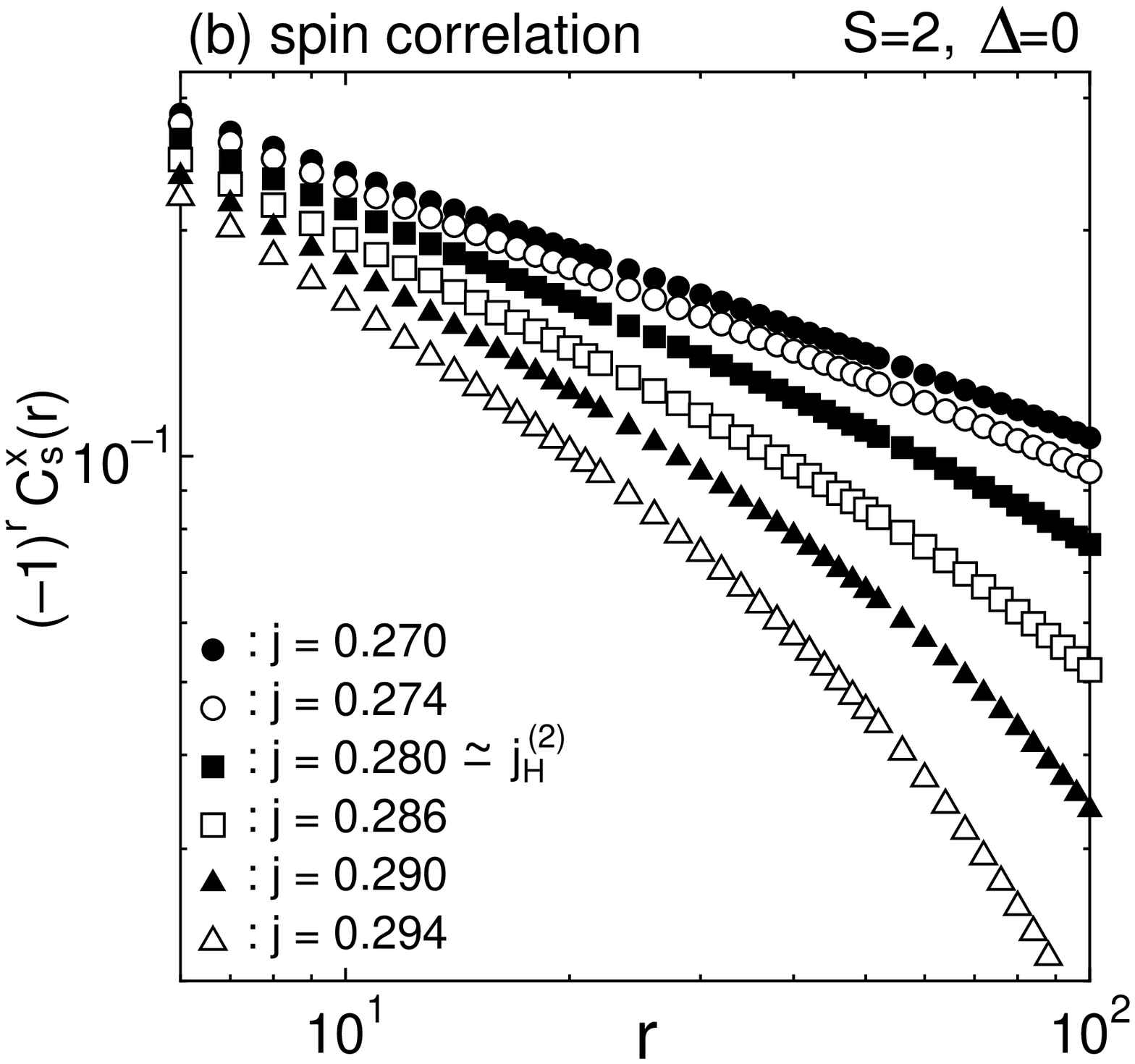}
\end{center}
\caption{Various correlation functions of the $S = 2$ chain 
for $\Delta = 0$ and for several typical values of $j$ 
around the SF-Haldane transition: 
(a) generalized string correlation function $C_{\rm str} (r)$;
(b) spin correlation function $(-1)^r C_s^x(r)$.
The number of kept states is $m = 220$.
The truncation errors are smaller than the symbols.
}
\label{fig:2SF-H}
\end{figure}

Let us first consider the SF-Haldane transition at $j = j_H^{(2)}$.
In Fig. \ref{fig:2SF-H}, we show the generalized string and spin correlation 
functions $C_{\rm str} (r)$ and $C_s^x (r)$ 
for $\Delta = 0$ and for several typical values of $j$ on $\log$-$\log$ plots.
It can be seen in Fig. \ref{fig:2SF-H} (a) that 
$C_{\rm str} (r)$ is bent upward for 
$j > j_H^{(2)} \simeq 0.280$ suggesting a finite LRO, 
while it shows a linear behavior for $j < j_H^{(2)}$ 
suggesting an algebraic decay.
Meanwhile, as shown in Fig. \ref{fig:2SF-H} (b), 
$C_s^x (r)$ decays exponentially for $j > j_H^{(2)}$ indicating a finite gap, 
whereas it decays algebraically for $j < j_H^{(2)}$.
We thus conclude that there occurs a phase transition at $j = j_H^{(2)}$ 
between the SF and the Haldane phases.
The critical value $j_H^{(2)}$ is estimated 
to be $j_H^{(2)} = 0.280 \pm 0.006$.
We estimate the critical points for various fixed $\Delta$ ($j$), 
and determine the phase boundary.
The transition line smoothly connects the point 
$j = j_H^{(2)}(\Delta = 0) = 0.280 \pm 0.006$ in the $XY$ case 
to the point of the case of no frustration, 
$\Delta = \Delta_H^{(2)}(j = 0) = 0.96 \pm 0.01$. 
(See the phase diagram shown in Fig. \ref{fig:2diag}.) 
The latter estimate is consistent with the previous estimate 
by Nomura and Kitazawa, $\Delta_H^{(2)}(j = 0) \simeq 0.966$.~\cite{No-Ki}

\begin{figure}
\begin{center}
\noindent
\leavevmode\epsfxsize=75mm
\epsfbox{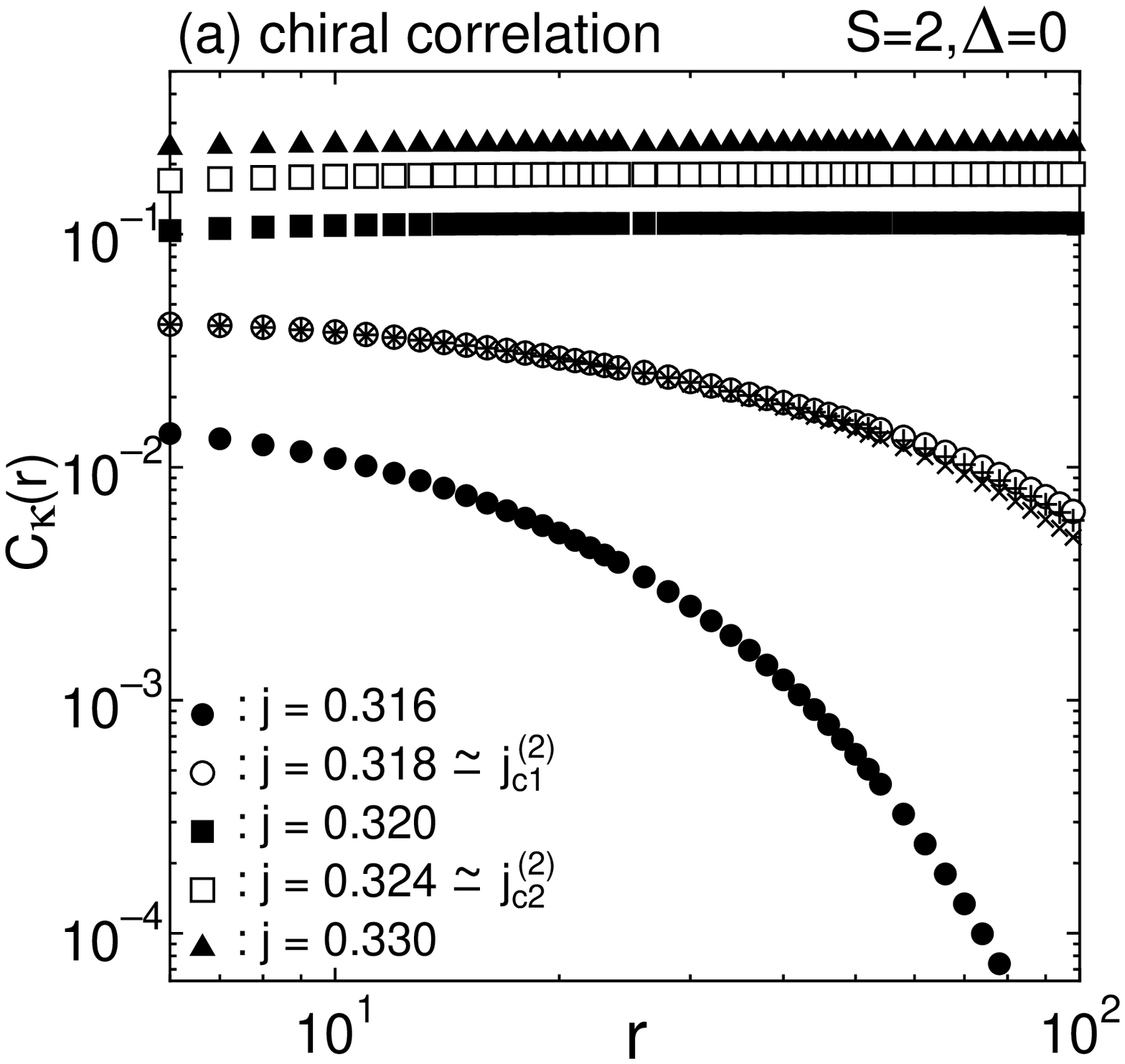}
%\end{center}
%\begin{center}
\noindent
\leavevmode\epsfxsize=75mm
\epsfbox{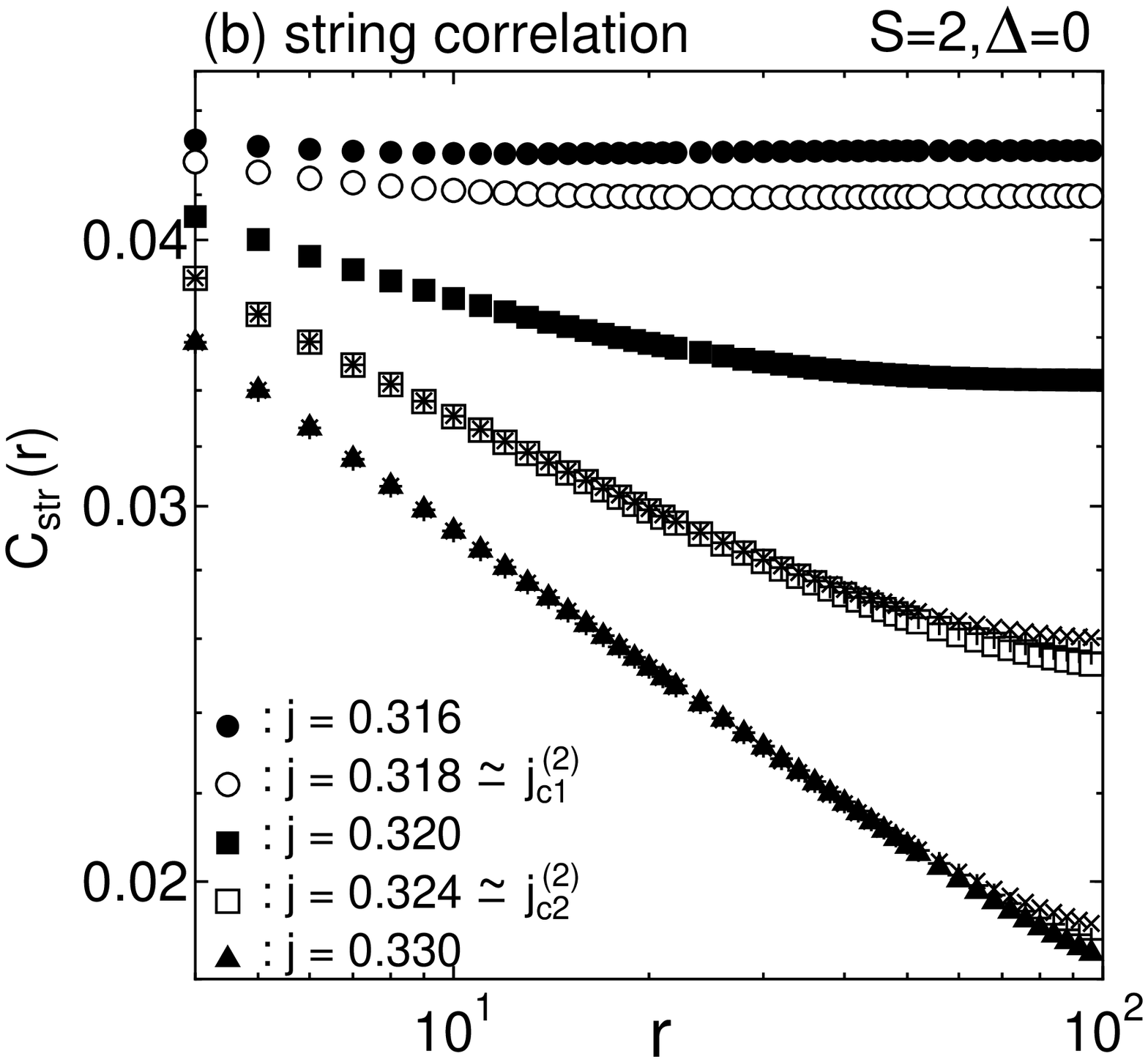}
\end{center}
\begin{center}
\noindent
\leavevmode\epsfxsize=75mm
\epsfbox{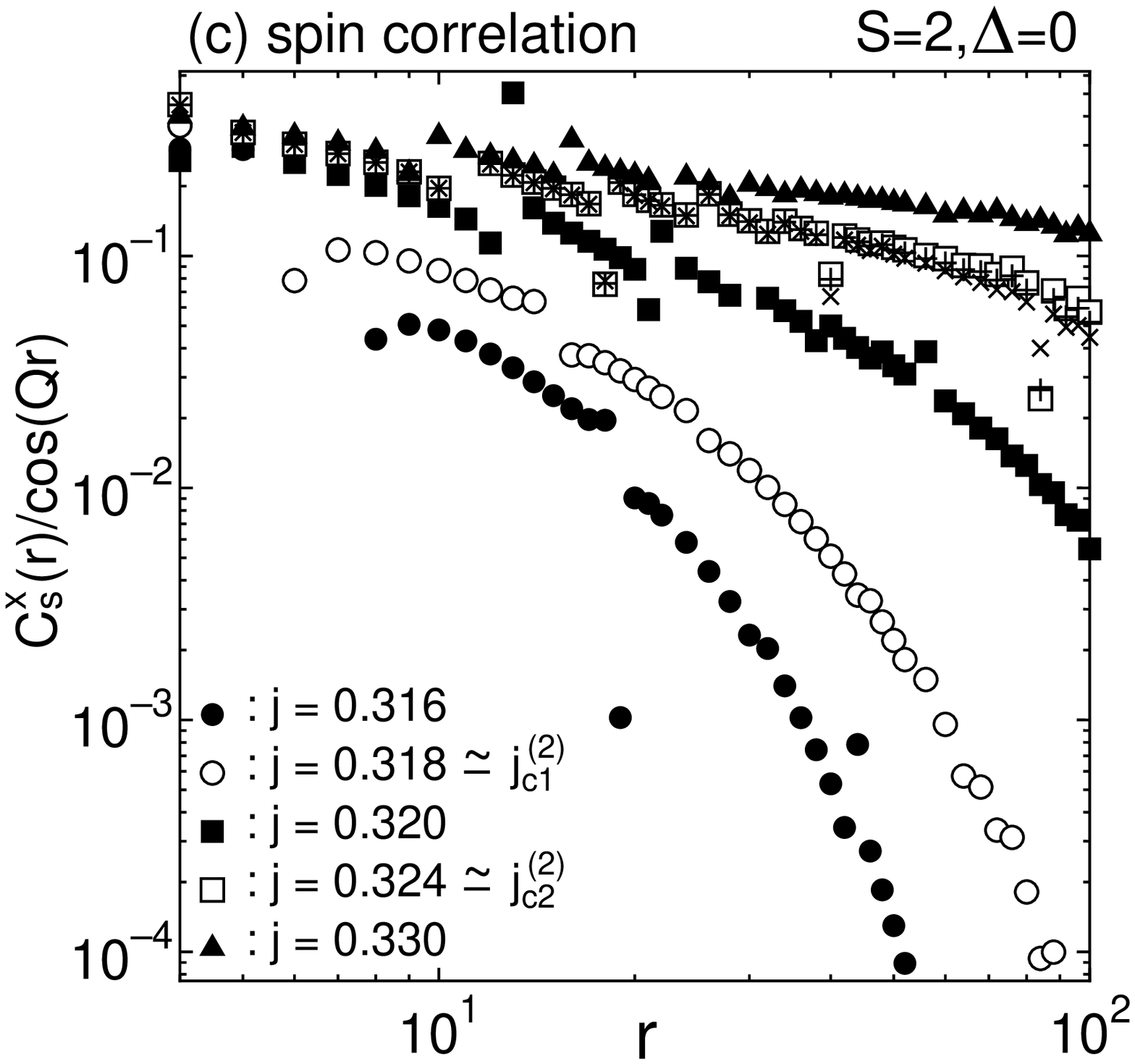}
\end{center}
\caption{Various correlation functions of the $S = 2$ chain 
for $\Delta = 0$ and for several typical values of $j$ 
around the Haldane-chiral transition: 
(a) chiral correlation function $C_\kappa (r)$;
(b) generalized string correlation function $C_{\rm str} (r)$;
(c) spin correlation function $C_{s}^x(r)$ divided by the oscillating factor 
$\cos (Qr)$. 
The number of kept states is $m = 260$.
To illustrate the $m$-dependence, we also 
indicate by crosses the data with $m = 220$ and $180$ 
for $j = 0.318$ in figure (a), $j = 0.324$ and $0.330$ in figure (b), 
and $j = 0.324$ in figure (c).
In other cases, the truncation errors are smaller than the symbols.
}
\label{fig:2H-C}
\end{figure}

Next, we consider the transition between the Haldane and the chiral phases.
Fig. \ref{fig:2H-C} (a) exhibits the chiral correlation function 
$C_\kappa (r)$ for $\Delta = 0$ and for several typical values of $j$.
As can clearly be seen from the figure, 
$C_\kappa (r)$ exhibits a finite LRO for $j > j_{c1}^{(2)} \simeq 0.318$ 
whereas it exhibits an exponential decay for $j < j_{c1}^{(2)}$.
We estimate the transition point where the chiral LRO sets in 
to be $j_{c1}^{(2)} = 0.318 \pm 0.001$.
As shown in Fig. \ref{fig:2H-C} (b), 
the generalized string correlation $C_{\rm str} (r)$ 
exhibits a finite LRO for $j < j_{c2}^{(2)} \simeq 0.324$ 
whereas it decays algebraically for $j > j_{c2}^{(2)}$.
Meanwhile, as shown in Fig. \ref{fig:2H-C} (c), 
the spin correlation function $C_s^x(r)$ divided by the leading oscillating 
factor $\cos(Qr)$ decays exponentially for $j < j_{c2}^{(2)}$ 
whereas it decays algebraically for $j > j_{c2}^{(2)}$.
From these observations, we estimate the transition point where 
the excitation spectrum becomes gapless and the generalized string order 
vanishes to be $j_{c2}^{(2)} = 0.324^{+0.006}_{-0.002}$.
Here the estimate of $j_{c2}^{(2)}$ is quite close to, 
but distinctly larger than that of $j_{c1}^{(2)} = 0.318 \pm 0.001$.
Indeed, the data of $j = 0.320$ in Fig. \ref{fig:2H-C} clearly show 
the existence of an intermediate phase, the gapped chiral phase, 
where the chiral and the generalized string LRO's coexist 
and the spin correlation decays exponentially.
Hence, we conclude that, as in the case of the $S = 1$ chain, 
the gapped chiral (chiral Haldane) phase exists also in the $S = 2$ chain 
in a very narrow but finite region ($j_{c1}^{(2)} < j < j_{c2}^{(2)}$) 
between the Haldane phase ($j < j_{c1}^{(2)}$) 
and the gapless chiral phase ($j > j_{c2}^{(2)}$).
Performing the same calculations for various fixed $\Delta$, 
we estimate the transition points 
$j_{c1}^{(2)}$ and $j_{c2}^{(2)}$.
The chiral Haldane phase is also found in a certain range of $j$ 
for $\Delta = 0.2$, $0.4$, $0.6$, and $0.8$.
By contrast, we can not confirm its existence 
for $\Delta > 0.8$ corresponding to $j \gsim 0.5$ 
within our numerical accuracy.
This is due to the large truncation error growing drastically 
with increasing $j$, 
which prevents us from precisely determining the transition points.

The obtained phase diagram of the $S = 2$ chain is shown 
in Fig. \ref{fig:2diag}.
The gapless chiral phase appears in a quite broad region of large $j$.
The boundary of the region with the chiral LRO, $j_{c1}^{(2)}$, 
rises steeply as $j$ increases from the point on the $XY$ line, 
$j_{c1}^{(2)}(\Delta = 0) \simeq 0.318$, 
tending to the Heisenberg line $\Delta = 1$.
We note that, although our estimates of the points 
where the chiral LRO sets in are very close to $\Delta = 1$ for $j \ge 0.6$, 
i.e., $\Delta_{c1}^{(2)} = 0.9975 \pm 0.0025$ 
for all $j = 0.6$, $0.7$, $0.8$, and $1.0$,  
the chiral LRO is not observed on the Heisenberg line $\Delta = 1$
where there no longer exists the two-fold discrete chiral degeneracy.

We finally refer to our numerical results 
on the possible double-Haldane (DH) phase.
The DH phase was first found in the large $j$ region of 
the frustrated $S = 1$ Heisenberg chain.~\cite{DH} 
In the DH phase, the next-nearest-neighbor coupling $J_2$ is dominant 
and the system can be regarded as two Haldane subchains 
coupled by the weak inter-subchain coupling $J_1$.
The DH phase is characterized by the absence of 
the string, spin, and chiral LRO's.~\cite{DHorder}
It has been shown that in the frustrated $S = 1$ $XXZ$ chain 
with $\Delta \gsim 0.95$, there occurs a first order phase transition 
between the Haldane and the DH phases.~\cite{HKKT,DH}
The string order parameter vanishes discontinuously at the transition point.
In our present study on the frustrated $S = 2$ Heisenberg chain, 
such a vanishing of the string order has not been observed: 
The generalized string correlation function $C_{\rm str}(r)$ 
in the Heisenberg case ($\Delta = 1$) exhibits a finite LRO 
for an entire region studied here, $0 \le j \le 1$, 
while both the spin and chiral correlation functions 
$C_s^x(r)$ and $C_\kappa(r)$ show an exponential decay 
for $0 \le j \le 1$ suggesting a gapfull excitation.
We note that, as $j$ increases, the extrapolated value 
$C_{\rm str}(r \to \infty)$ decreases rapidly around $j \simeq 0.4$.
Although the rapid drop of the string LRO might be a sign of 
a phase transition from the Haldane phase 
to a new intermediate phase between the Haldane and the DH phases 
as suggested in Ref. \onlinecite{Roth}, 
the situation here remains largely unclear.
Further work is required to clarify the details of the transition 
at and near $\Delta = 1$ and 
the possible existence of the DH phase for $j \gsim 1$.

\section{$S$-dependence}

Based on our present results for $S = 1/2$, $3/2$, and $2$ and 
our previous result for $S = 1$,~\cite{HKKT} 
we now discuss how the ground-state phase diagrams change 
as $S$ increases from $S = 1/2$ toward the classical limit $S \to \infty$.
We are interested particularly in the way how the classical limit is achieved 
from the quantum phases analyzed above.

By comparing the obtained phase diagrams for $S = 1/2$, $1$, $3/2$, and $2$, 
we deduce several features of the phase diagrams.
The first is about the fate of the dimer and the Haldane phases. 
These phases are pure quantum ones in the sense that 
singlet spin pairs play an essential role in stabilizing them.
Hence, it is natural to expect that 
the regions of these quantum phases become narrower as $S$ gets larger.
This feature can clearly be seen in the phase diagrams 
obtained for $1/2 \le S \le 2$: 
The dimer phase for $S = 3/2$ is narrower than that for $S = 1/2$, 
while the Haldane phase for $S = 2$ is narrower than that for $S = 1$.
It thus seems reasonable to expect that 
the dimer and the Haldane phases continue to become narrower for $S > 2$, 
and eventually vanish in the classical limit $S \to \infty$.

The second feature concerns the chiral ordered phases.
Recently, the $S$-dependence of the boundary of the chiral ordered phases 
was examined via the field-theoretical large-$S$ approach.~\cite{Kole}
It was shown there that, as $S$ increased, 
the region of the gapless chiral phase converged smoothly 
to that of the helical ordered phase in the classical limit $S \to \infty$, 
whereas the gapped chiral phase vanished asymptotically.
This feature of the chiral phases might be understood intuitively by 
considering the role of quantum fluctuations in the symmetry breaking.
In the helical ordered state with a finite magnetic LRO 
realized in the classical limit $S \to \infty$, 
both the discrete $Z_2$ parity and the continuous $U(1)$ spin symmetries 
are broken spontaneously.
In the gapless chiral phase, quantum fluctuations marginally recover 
the continuous $U(1)$ symmetry yielding the quasi-long-range spin order, 
with keeping the discrete $Z_2$ parity symmetry broken.
In this sense, the gapless chiral phases can be regarded as 
a quantum remnant of the classical helical phase.
By contrast, the gapped chiral phase should be regarded as 
a pure quantum phase since it exhibits a topological LRO, i.e., 
the string LRO, in the chiral Haldane phase for integer $S$.
(In the chiral dimer phase for half-odd integer $S$, 
if any, the topological LRO is the dimer LRO, 
although we donot find evidence of such a phase in the present study.)
We therefore expect that the gapped chiral phase 
should vanish in the classical limit $S \to \infty$.
The observed behaviors of the chiral phases in the obtained phase diagrams 
are consistent with the above expectation.
With increasing $S$, the chiral transition line $j_{c1}^{(S)}(\Delta)$ 
approaches that of the helical ordered phase, $j = 1/4$.
Meanwhile, the region of the chiral Haldane phase found for integer $S$ 
shrinks as $S$ increases.
As an example, we list in Table \ref{tab:j-S} the estimated values of 
$j_{c1}^{(S)}$ and $j_{c2}^{(S)}$ in the $XY$ case ($\Delta = 0$) 
and their ratio $(j_{c2}^{(S)}-j_{c1}^{(S)})/j_{c1}^{(S)}$ for integer $S$, 
which is a measure of the relative stability of the chiral Haldane phase.

The third feature concerns the SF phase.
One may naturally expect that, as $S$ increases, 
the region of the SF phase converges to that of the N\'{e}el phase 
in the classical limit $S \to \infty$.
Somewhat unexpectedly, in the obtained phase diagrams, the SF phase grows 
as $S$ becomes larger, {\it exceeding} the classical phase boundary 
of the N\'{e}el phase, $j = 1/4$.
(See the phase diagrams and Table \ref{tab:j-S} in which 
 the estimates of $j_d^{(S)}$ and $j_H^{(S)}$ in the $XY$ case are listed.)
We consider that this ^^ ^^ overshooting" is due to the rapid shrink of 
the dimer and the Haldane phases 
and that the boundary of the SF phase eventually ^^ ^^ turns back" 
converging to the classical phase boundary $j = 1/4$ for large enough $S$.
In order to confirm this conjecture, 
the analysis of larger $S$ is needed, 
which is beyond the scope of the present work.

\begin{figure}
\begin{center}
\noindent
\leavevmode\epsfxsize=75mm
\epsfbox{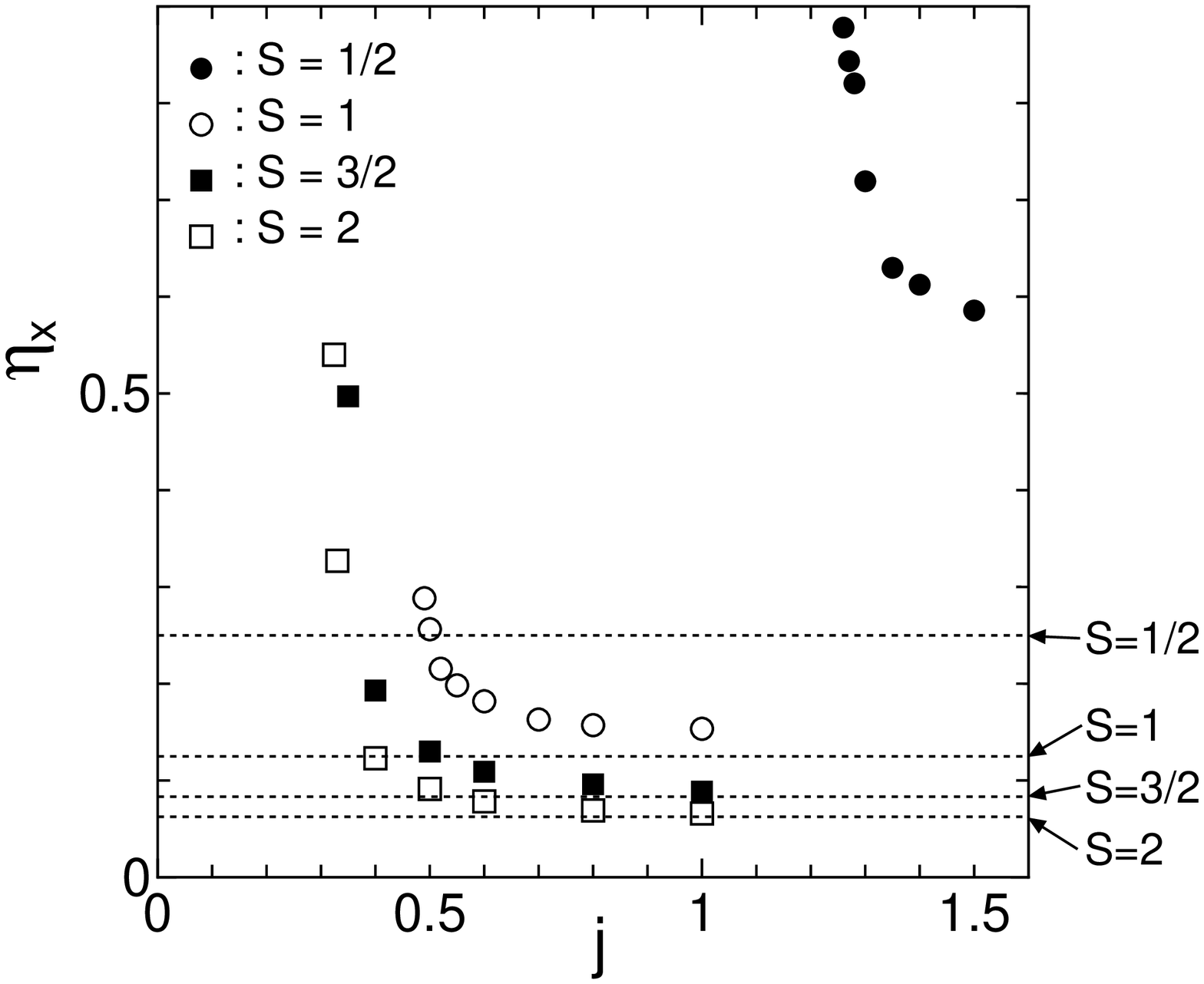}
\end{center}
\caption{The estimated decay exponent $\eta_x$ as a function of $j$ 
for the $S = 1/2, 1, 3/2$, and $2$ $XY$ chains.
The dotted lines represent the prediction 
of the bosonization analysis~\cite{Ne,LJA} at $j \to \infty$, 
$\eta_x = 1/(8S)$.
}
\label{fig:eta-S}
\end{figure}

Finally, we wish to compare our numerical result on the decay exponent 
of the spin correlation 
with the prediction from the bosonization analysis.~\cite{Ne,LJA}
Applying the bosonization technique and the mean-field approximation, 
the authors of Refs. \onlinecite{Ne} and \onlinecite{LJA} predicted that 
for large enough $j$ the spin correlation function $C_s^x (r)$ 
in the $XY$ case should exhibit the asymptotic behavior, 
\begin{equation}
C_s^x (r) = A \cos(Qr) r^{-\eta_x},  \label{eq:eta}
\end{equation}
where $A$ is a numerical constant and 
the exponent $\eta_x$ is given by $\eta_x = 1/(8S)$.
We fit our numerical data 
for the $S = 1/2$, $3/2$, and $2$ $XY$ chains with Eq. (\ref{eq:eta}), 
taking $A$, $Q$, and $\eta_x$ as fitting parameters.
Resulting estimates of the decay exponent $\eta_x$ 
are shown in Fig. \ref{fig:eta-S} together with our previous estimate 
for the $S = 1$ $XY$ chain.~\cite{HKKT}
As can be seen from Fig. \ref{fig:eta-S}, 
the estimates of $\eta_x$ decrease monotonically as $j$ increases.
For $S \ge 1$, the asymptotic $j \to \infty$ value 
has almost been attained at around $j \simeq 1$, 
which is in good agreement 
with the predicted value $1/(8S)$.
For the $S = 1/2$ case, on the other hand, 
we can not reach the asymptotic $j \to \infty$ regime: 
The estimated $\eta_x$ still continues 
to decrease even around $j \simeq 1.5$, 
the largest $j$ value for which we can get reliable data.
The estimated $\eta_x$, however, shows a tendency to further decrease 
toward the predicted value $\eta_x = 1/4$ as $j$ increases, 
which is consistent with the result of the bosonization study.
Thus, our result can be considered as a numerical support 
of the bosonization analysis.

\section{Summary}

\begin{figure}
\begin{center}
\noindent
\leavevmode\epsfxsize=75mm
\epsfbox{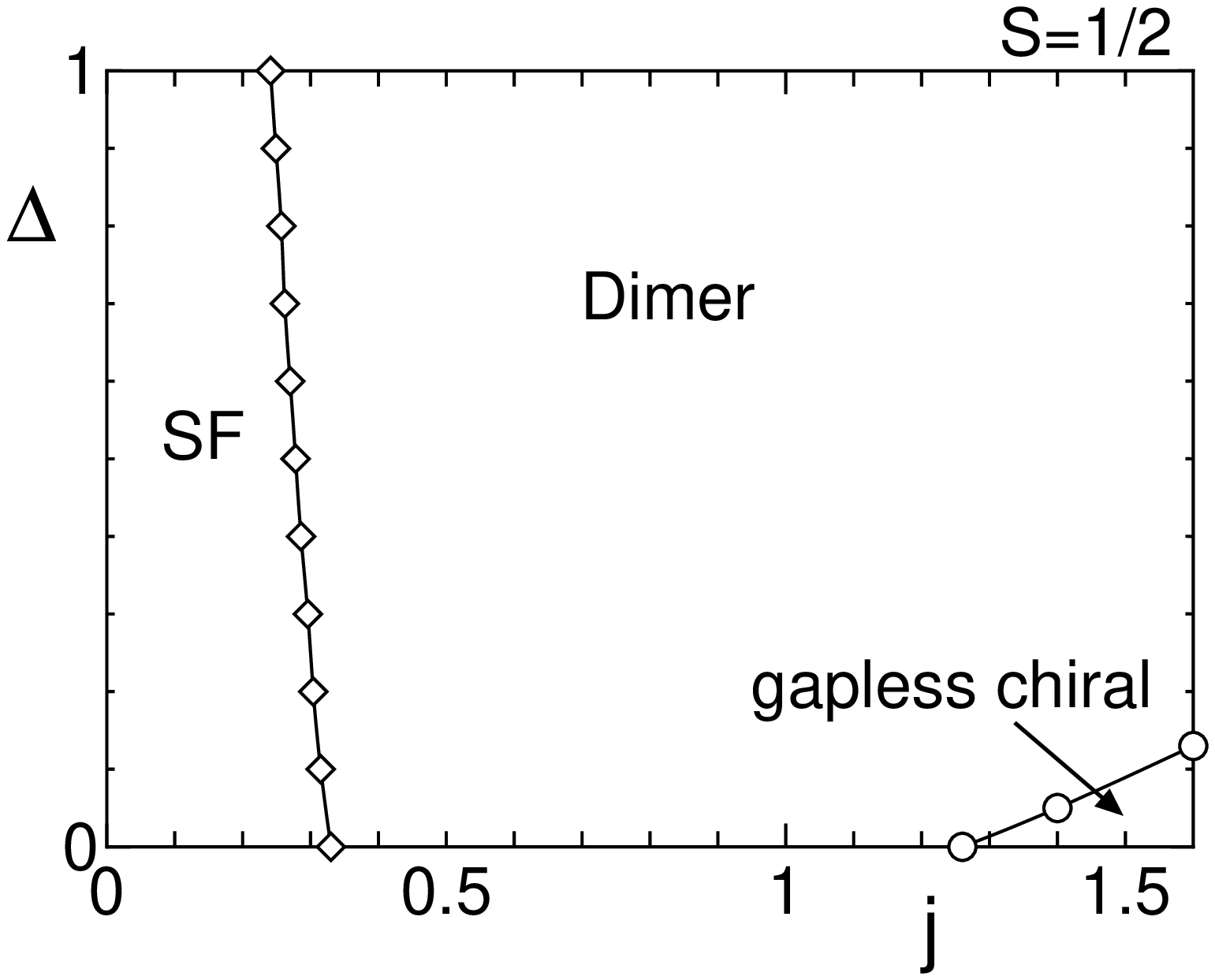}
\end{center}
\caption{The ground-state phase diagram of the $S = 1/2$ chain, 
where $j$ and $\Delta$ denote the ratio $J_2/J_1$ and 
the exchange anisotropy, respectively, defined in Eq. (\ref{eq:Ham}).
The diamonds and circles represent the transition points 
$j_d^{(1/2)}$ and $j_{c1}^{(1/2)}$.
}
\label{fig:1hdiag}
\end{figure}

\begin{figure}
\begin{center}
\noindent
\leavevmode\epsfxsize=75mm
\epsfbox{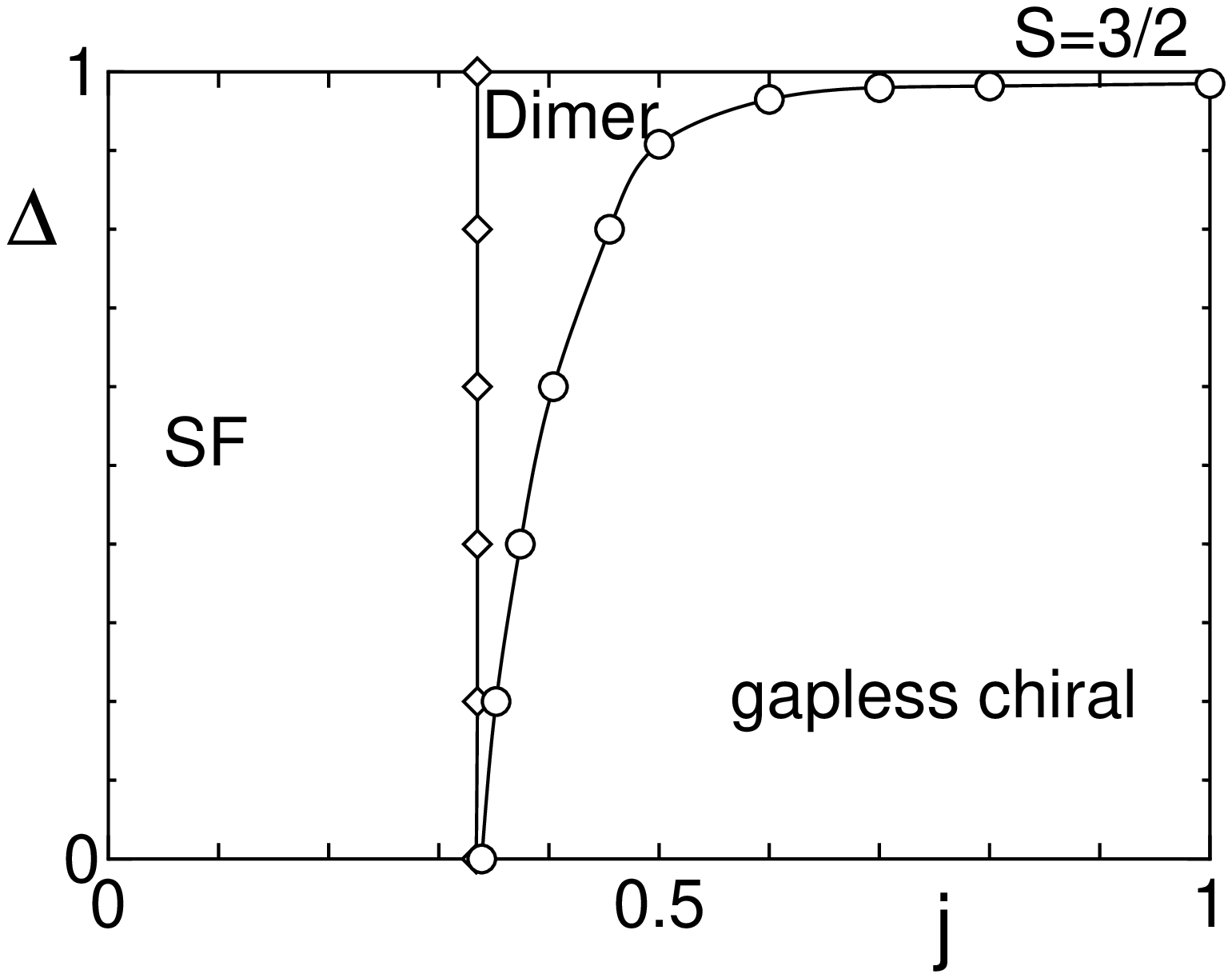}
\end{center}
\caption{The ground-state phase diagram of the $S = 3/2$ chain, 
where $j$ and $\Delta$ denote the ratio $J_2/J_1$ and 
the exchange anisotropy, respectively, defined in Eq. (\ref{eq:Ham}).
The diamonds and circles represent the transition points 
$j_d^{(3/2)}$ and $j_{c1}^{(3/2)}$.
}
\label{fig:3hdiag}
\end{figure}

\begin{figure}
\begin{center}
\noindent
\leavevmode\epsfxsize=75mm
\epsfbox{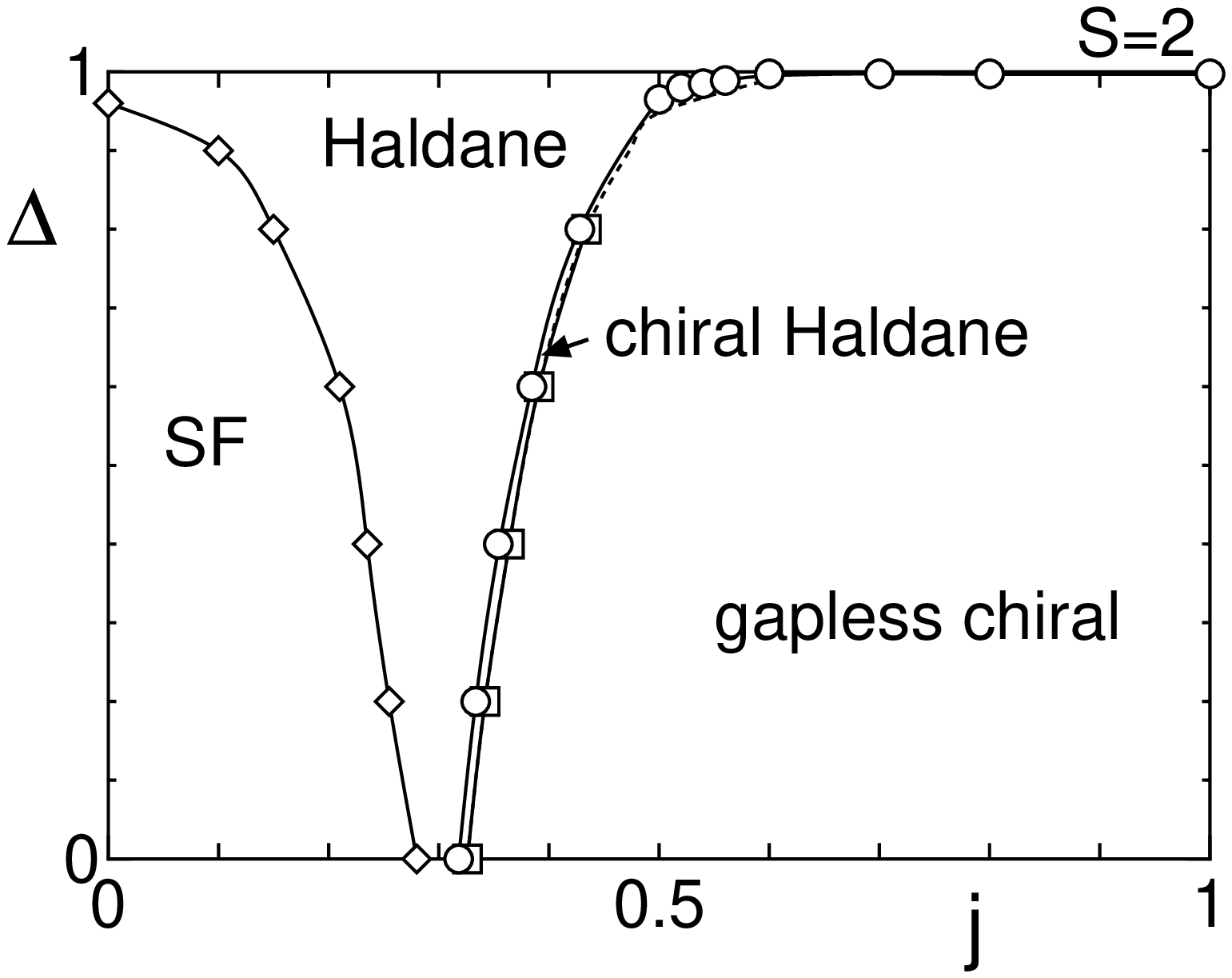}
\end{center}
\caption{The ground-state phase diagram of the $S = 2$ chain, 
where $j$ and $\Delta$ denote the ratio $J_2/J_1$ and 
the exchange anisotropy, respectively, defined in Eq. (\ref{eq:Ham}).
The diamonds, circles, and squares represent the transition points 
$j_d^{(2)}$, $j_{c1}^{(2)}$, and $j_{c2}^{(2)}$, respectively.
}
\label{fig:2diag}
\end{figure}

In the present work, we have studied the ground-state properties 
of the frustrated spin-$S$ $XXZ$ chains (\ref{eq:Ham}), 
especially paying attention to the chiral ordered phases 
in which only the chirality exhibits a finite LRO 
without the standard magnetic LRO.
We have used the infinite-system DMRG method 
to calculate the correlation functions 
associated with the spin, chiral, dimer, and string order parameters.
By analyzing the long-distance behavior of the correlation functions, 
we have determined the ground-state phase diagrams 
of the $S = 1/2$, $3/2$, and $2$ chains 
for $0 \le \Delta \le 1$ and $j \ge 0$ 
(Figs. \ref{fig:1hdiag}, \ref{fig:3hdiag}, and \ref{fig:2diag}).

By comparing the obtained results with our previous result 
for the $S = 1$ chain,  we reach the following picture of the 
ground-state phase diagram.
In the integer $S$ chains ($S = 1$ and $2$), 
the phase diagram consists of four different phases, i.e., 
the SF, Haldane, gapless chiral, and gapped chiral (chiral Haldane) phases.
In the half-odd integer $S$ chains ($S = 1/2$ and $3/2$), 
on the other hand, we have found three phases, i.e., 
the SF, dimer, and gapless chiral phases.
It thus turns out that the gapless chiral phase 
appears for general $S \ge 1/2$.
For integer $S$, the gapped chiral (chiral Haldane) phase 
exists in a narrow region 
between the Haldane and the gapless chiral phases.
For half-odd integer $S$, by contrast, 
the gapped chiral (chiral dimer) phase 
has not been identified within our numerical precision: 
The rather large truncation error of the DMRG calculation 
prevents us from verifying whether the chiral dimer phase exists or not.
Our results suggests, however, that the chiral dimer phase, 
if it ever exists, appears in a narrow region between 
the dimer and the gapless chiral phase.
Further work will be required to solve the problem.

We have also discussed the $S$-dependence of the phase diagrams.
The obtained phase diagrams indicate that, 
as $S$ increases toward the classical limit $S \to \infty$, 
the region of the gapless chiral phase 
converge smoothly toward that of the classical helical phase, 
while the pure quantum phases, i.e., the dimer, Haldane, and 
gapped chiral phases, become narrower and 
eventually vanish in the $S \to \infty$ limit.
The prediction from the bosonization study 
that the decay exponent of the spin correlation 
in the $XY$ case ($\Delta = 0$) 
takes a value $\eta_x = 1/(8S)$ in the $j \to \infty$ limit 
has been verified.

Finally, we wish to touch upon the possible experimental realization 
of the gapless chiral phase in a quasi-one-dimensional compound 
${\rm CaV}_2{\rm O}_4$. 
This material is expected to be described by 
the frustrated $S = 1$ chain (\ref{eq:Ham}) 
where the AF next-nearest-neighbor coupling $J_2$ 
is comparable to the AF nearest-neighbor coupling $J_1$.
Recently, Kikuchi made measurements on magnetic susceptibility 
and on $^{51}{\rm V}$ NMR, and 
showed that the system had gapless excitations 
above the ground state.~\cite{Kiku}
For the frustrated $S = 1$ chain (\ref{eq:Ham}), 
theoretical studies indicate that 
there is no gapless phase for $j \ge 0$ and $\Delta \ge 0$ 
except for the gapless chiral phase.
Hence, ${\rm CaV}_2{\rm O}_4$ might be a promising candidate 
for the realization of the chiral ordered phase.
For the future, it might be interesting to calculate thermodynamic 
properties at finite temperatures to compare them with 
the experimental data on this compound.

\acknowledgements
We thank H. Kikuchi for fruitful discussion.
Numerical calculations were carried out in part at the Yukawa
Institute Computer Facility, Kyoto University.
T.H. was supported by the Japan Society 
for the Promotion of Science for Young Scientists.

\end{multicols}

\newpage
\begin{table}
\caption{The estimates of $j_d^{(S)}$, $j_H^{(S)}$, 
$j_{c1}^{(S)}$, and $j_{c2}^{(S)}$ in the $XY$ case ($\Delta = 0$) 
for half-odd integer $S$ ($S = 1/2$ and $3/2$) and 
for integer $S$ ($S = 1$ and $2$).
The ratio $(j_{c2}^{(S)}-j_{c1}^{(S)})/j_{c1}^{(S)}$ for integer $S$ 
is also listed.
}
\label{tab:j-S}
\begin{tabular}{cccccc}
$S$ & $j_d^{(S)}$ & $j_H^{(S)}$ & $j_{c1}^{(S)}$ & $j_{c2}^{(S)}$ & 
$(j_{c2}^{(S)}-j_{c1}^{(S)})/j_{c1}^{(S)}$  \\ \hline
$1/2$ & $0.33^a$          & &$1.26^{+0.01}_{-0.03}$ & $1.26 \pm 0.01$  &  \\
$3/2$ & $0.334 \pm 0.004$ & &$0.339 \pm 0.001$ & $0.340^{+0.004}_{-0.002}$ 
&         \\ \hline
$1$ & & $0$         & $0.473 \pm 0.001$ & $0.490^{+0.010}_{-0.005}$ 
& $0.036$ \\
$2$ & & $0.280 \pm 0.006$ & $0.318 \pm 0.001$ & $0.324^{+0.006}_{-0.002}$ 
& $0.019$ 
\end{tabular}
$^a$Ref. \onlinecite{Oka-Nomu}. \\
The data for $S = 1$ are from Ref. \onlinecite{HKKT}.
\end{table}

\newpage

\begin{thebibliography}{99}
\bibitem{Cdim}
To be more precise, the dimer correlation in the SF phase 
still decays even at $r = 100$ although the decay rate 
is extremely small and becomes smaller as $r$ increases.
Consequently, the value of $C_{\rm dim}^x (r=100)$ in the SF phase 
slightly overestimates the extrapolated value $C_{\rm dim}^x (r \to \infty)$.
\bibitem{apparent}
More precisely, we believe that the effect of open boundaries 
is also essential to the apparent LRO of $C_{\rm dim}^x(r)$. 
Generally speaking, at least when $J_2$ is not so large, 
the correlation of a spin pair on the edge of an open chain 
is stronger than that in the bulk of the chain, 
while the correlation of a spin pair next to the edge is weaker.
Such a ^^ ^^ strong-weak" pattern propagates into the bulk of the chain.
When the system is in the gapless phases, 
such deviation of the spin correlation is expected to 
decay algebraically as the edge of the chain goes away. 
We thus consider that we should observe an algebraic decay of 
the boundary effect if we do the DMRG calculation 
for the gapless phases with $m \to \infty$.
In contrast to this expectation, 
our present calculation with finite $m$ yields 
an apparent dimer LRO in the gapless phases, 
which should be attributed to 
the truncation error associated with the boundary effect.
\bibitem{DHorder}
Since the system in the DH phase can be described approximately 
by the direct product of two Haldane subchains, 
one may expect the string correlations on each subchain 
to be the order parameter of the phase.
Contrary to this naive expectation, however, 
the subchain string correlation in the DH phase does not exhibit 
a finite LRO but decays exponentially.
(See Ref. \onlinecite{DH}.)
\end{thebibliography}
\end{document}